\documentclass{aa}

\usepackage{psfig}
\usepackage{amssymb}
\newcommand{\hii}{H~{\sc ii}}
\newcommand{\Hei}{He~{\sc i}}
\newcommand{\Heii}{He~{\sc ii}}
\newcommand{\hei}{He~{\sc i} $\lambda$4471 }
\newcommand{\heii}{He~{\sc ii}  $\lambda$4542 }

\newcommand{\teff}{\ifmmode T_{\rm eff} \else $T_{\mathrm{eff}}$\fi}
\newcommand{\logg}{\ifmmode \log g \else $\log g$\fi}
\newcommand{\msun}{\ifmmode M_{\odot} \else M$_{\odot}$\fi}
\newcommand{\rsun}{\ifmmode R_{\odot} \else R$_{\odot}$\fi}
\newcommand{\zsun}{\ifmmode Z_{\odot} \else Z$_{\odot}$\fi}
\newcommand{\lsun}{\ifmmode L_{\odot} \else L$_{\odot}$\fi}

\newcommand{\mdot}{\ifmmode \dot{M} \else $\dot{M}$\fi}
\newcommand{\vesc}{\ifmmode v_{\rm esc} \else $v_{\rm esc}$\fi}
\newcommand{\vinf}{\ifmmode v_{\infty} \else $v_{\infty}$\fi}
\newcommand{\vturb}{\ifmmode v_{\rm turb} \else $v_{\rm turb}$\fi}
\newcommand{\mv}{\ifmmode M_{\rm{V}} \else M$_{\rm{V}}$\fi}
\newcommand{\lL}{\ifmmode \log \frac{L}{L_{\odot}} \else $\log \frac{L}{L_{\odot}}$\fi}
\newcommand{\qo}{\ifmmode \log q_{0} \else $\log q_{0}$\fi}
\newcommand{\qi}{\ifmmode \log q_{1} \else $\log q_{1}$\fi}
\newcommand{\Qo}{\ifmmode \log Q_{0} \else $\log Q_{0}$\fi}
\newcommand{\Qi}{\ifmmode \log Q_{1} \else $\log Q_{1}$\fi}

\newcommand{\pv}{P~{\sc v} $\lambda\lambda$1118,1128 }

\newcommand{\ov}{O~{\sc v}  $\lambda$1371 }

\newcommand{\kms}{km s$^{-1}$}

\def\aap{A\&A}
\def\aaps{A\&AS}

\def\apj{ApJ}

\def\apjs{ApJS}
\def\mnras{MNRAS}

\begin{document}

%%%%%%%%%%%%%%%%%%%%%%%%%%%%%%%%%%%%%%%%%%%%%%%%%%%%%%%%%%%%%%%%%%%%%%%%
%\begin{tabular}

%\thesaurus{} 

\title{A new calibration of stellar parameters of Galactic O stars}
%\thanks{}

\author{Fabrice Martins 
           \inst{1,2}\thanks{\textit{Present address}: Max-Planck Institut 
f$\ddot{\rm u}$r Extraterrestrische Physik, Postfach-1312, D-85741, Garching, Germany} 
        \and Daniel Schaerer 
           \inst{1,2}
        \and D. John Hillier
           \inst{3}}

\offprints{F. Martins, martins@mpe.mpg.de \inst{1}}

 \institute{Observatoire de Gen\`eve, 51 Chemin des Maillettes, CH-1290 Sauverny, 
 Switzerland
 \and
 Laboratoire d'Astrophysique, Observatoire Midi-Pyr\'en\'ees
 14 Av.  E. Belin, F-31400 Toulouse, France
 \and 
 Department of Physics and Astronomy, University of Pittsburgh, 3941
 O'Hara Street, Pittsburgh, PA 15260, USA
}

\date{Received 18 November 2004 / Accepted 14 March 2005}

\titlerunning{Calibration of O star parameters}
%\authorrunning{}

%%%%%%%%%%%%%%%%%%%%%%%%%%%%%%%%%%%%%%%%%%%%%%%%%%%%%%%%%%%%%%%%%%%%%%%%
\abstract{
 We present new calibrations of stellar parameters of O stars at solar 
metallicity taking non-LTE, wind, and line-blanketing effects into account. 
Gravities and absolute visual magnitudes are derived from 
results of recent spectroscopic analyses.
Two types of effective temperature scales are derived: one from a compilation 
based on recent spectroscopic studies of a sample of massive stars 
-- the ``observational scale'' --
and the other from direct interpolations on a grid 
of non-LTE spherically extended line-blanketed models computed with the 
code CMFGEN (Hillier \& Miller \cite{hm98}) -- the ``theoretical scale''. 
These \teff\ scales are then further used together with the grid of models to calibrate 
other parameters
(bolometric correction, luminosity, radius, spectroscopic 
mass and ionising fluxes) as a function of spectral type and luminosity class.
Compared to the earlier calibrations of Vacca et al.\ (\cite{vacca})
the main results are:

\begin{itemize}

\item The effective temperature 
scales of dwarfs, giants and supergiants are cooler by 2000 to 8000 K, 
the theoretical 
scale being slightly cooler than the observational one. The reduction 
is the largest for the earliest spectral types and for supergiants.

\item Bolometric corrections as a function of \teff\ are reduced by 0.1 mag
due to line blanketing which redistributes part of the UV flux in the optical range. 
For a given spectral type the reduction of $BC$ is 
larger for early types and for supergiants. Typically $BC$s derived using 
the theoretical \teff\ scale are 0.40 to 0.60 mag lower than that of 
Vacca et al.\ (\cite{vacca}), whereas the differences using the 
observational \teff\ scale are somewhat smaller.

\item Luminosities are reduced by 0.20 to 0.35 dex for dwarfs, 
by $\sim$ 0.25 for all giants and by 0.25 to 0.35 dex for supergiants.
The reduction is essentially the same for both \teff\ scales. It is independent 
of spectral type for giants and supergiants and is slightly larger for late type 
than for early type dwarfs.

\item Lyman continuum fluxes are reduced. 
Our theoretical values for the hydrogen ionising photon fluxes for dwarfs 
are 0.20 to 0.80 dex lower than those of 
Vacca et al.\ (\cite{vacca}), the difference being larger at late spectral 
types. For giants the reduction is of 0.25 to 0.55 dex, while for 
supergiants it is of 0.30 to 0.55 dex. Using the observational \teff\ 
scale leads to smaller reductions at late spectral types.

\end{itemize}

The present results should represent a significant improvement over previous calibrations, 
given the detailed treatment of non-LTE line-blanketing in the expanding atmospheres 
of massive stars. 

}

\maketitle

%%%%%%%%%%%%%%%%%%%%%%%%%%%%%%%%%%%%%%%%%%%%%%%%%%%%%%%%%%%%%%%%%%%%%%%%
%----------------------------------------------------------------------
%
%          Introduction
%
%----------------------------------------------------------------------
\section{Introduction}
\label{intro}

Despite their paucity, massive stars play a crucial role in several
fields of astrophysics: they enrich the interstellar medium in heavy
elements; they create \hii\ regions; they release huge quantities of
mechanical energy through their winds; they explode as supernovae;
they are possibly at the origin of Gamma-Ray Bursts; and the first
massive stars may have reionised the early Universe at redshift beyond 
6. Hence, a good knowledge of their properties is crucial and requires 
the development of both evolutionary and atmosphere models.

Three main ingredients have to be included in massive stars
atmospheres: a full non-LTE treatment since radiative processes 
are dominant over thermal processes (e.g. Auer \& Mihalas \cite{am72}); 
spherical expansion due to the stellar wind and the related velocity fields 
(Hamann \cite{hamann86}, Hillier \cite{hil87a}, \cite{hil87b}, 
Gabler et al.\ \cite{gabler89}, Kudritzki \cite{kud92}, Najarro et al.\ 
\cite{paco96}); 
and line-blanketing to take into account the effects of metals on 
the atmospheric structure and emergent spectrum (Abbott \& Hummer 
\cite{ah85}, Schaerer \& Schmutz \cite{ss94}). The two former 
ingredients were the first to be included in the models, but it is 
only recently that line-blanketing has been handled reliably. The 
main reason is the complexity of the problem which has to be solved 
when thousands of level populations from metals have to be computed 
through the resolution of statistical equilibrium and radiative transfer 
equations in an expanding medium. Various solutions have been developed 
to include line-blanketing: opacity sampling method in the code WM-BASIC 
(Pauldrach et al.\ \cite{wmbasic01}), approximate method to estimate 
line-blocking and blanketing in FASTWIND 
(Santolaya-Rey et al.\ \cite{fast97}, Puls et al.\ \cite{puls05}), 
opacity distribution functions in TLUSTY (Hubeny \& Lanz \cite{hl95}, 
Lanz \& Hubeny \cite{ostar2002}), 
or comoving frame calculations using super-levels in CMFGEN (Hillier \& 
Miller \cite{hm98}). Each method and code has its advantages and disadvantages: 
WM-BASIC makes a treatment in the observer's frame uising a Sobolev plus continuum 
approximation in the solution of the rate equations and does not include line 
broadening terms, but makes a complete hydrodynamical calculation of the atmosphere 
structure; FASTWIND is designed for fast computations but makes only an 
approximate treatment of opacities from metals for which no line profile is 
(yet) predicted; TLUSTY makes a very detailed calculation of the NLTE rate 
equations but is limited to plane-parallel geometry; finally, CMFGEN solves the NLTE 
rate equations in the comoving frame but uses super-levels. 
%(although this 
%approximation can be easily dropped provided the computational resources are 
%available). 

The effects of line-blanketing on atmosphere models lead to quantitative 
modifications of the stellar and wind properties of massive stars in 
general and O stars in particular. The most studied effect is the 
reduction of the effective temperature scale (Martins et al.\ 
\cite{teffscale}, Crowther et al.\ \cite{paul02}, Herrero et al.\ 
\cite{hpn02}, Bianchi \& Garcia \cite{bg02}, Repolust et al.\ \cite{repolust}, 
Massey et al.\ \cite{massey04}). 
Indeed, the increased 
number of diffusions in the inner atmosphere due to metallic line opacities
implies a heating of the 
deeper layers (backwarming effect) and consequently a higher ionisation 
which shifts the relation between effective temperature and spectral 
type. The reduction can be as high as 7000 K for extreme supergiants 
(Crowther et al.\ \cite{paul02}). Such a change of the effective 
temperatures implies lower luminosities and lower ionising fluxes, which 
is crucial for studies involving \hii\ regions and star forming regions. 

A good knowledge of the effective temperature scale of O stars is 
fundamental since \teff\ cannot be derived from optical photometry: 
the visual spectrum of O stars is in the Rayleight-Jeans part 
of the distribution and is thus almost insensitive to \teff. 
Several \teff\ scales have been proposed in the past 
(Conti \cite{conti73}, Schmidt-Kaler \cite{sk82}) 
based on models without winds and metals, the most recent one being 
that of Vacca et al.\ (\cite{vacca}). As these ingredients are now 
available in models, revisions of such calibrations are possible. Once 
obtained, they can be used to calibrate other important parameters such 
as luminosities and ionising fluxes.

Two main approaches lend themselves to derive a temperatures scale as a function
of spectral type (hereafter ST) and luminosity class (LC):
{\em 1)} the determination of average \teff's from an observed sample of O stars, or
{\em 2)} the determination of \teff's from the comparison of extended model atmosphere
grids with the observed mean properties defining the spectral types, i.e.\
\Hei/\Heii\ spectral line ratios for O stars.
Both methods, hereafter referred to as the ``observational scale'' {\em (1)} and
the ``theoretical scale'' {\em (2)} \footnote{Note that both scales rely on 
model atmospheres: one is related to observations of a sample of ``real'' massive stars 
analysed through spectroscopic with model atmospheres (hence the name 
``observational''), whereas the other one
is only based on the model grid without direct link to existing stars 
(hence the name ``theoretical'').}, are employed 
in the present work.
Their advantages and drawbacks are roughly the following.
Obviously the establishment of an ``observational scale'' relies on a sufficient
and representative number of individual stars of all spectral types
and luminosity classes, which have been analysed with state-of-the-art
non-LTE line blanketed model atmospheres with stellar winds.
Several such studies
have been pursued in the last years (Crowther et al.\ \cite{paul02}, Herrero et al.\ 
\cite{hpn02}, Hillier et al.\ \cite{hil03}, Bouret et al.\ \cite{jc03}, 
Repolust et al.\ \cite{repolust}, 
Markova et al.\ \cite{mark04}, Evans et al.\ \cite{evans04}, Martins et al.\ 
\cite{n81}, \cite{wwgal}). However, 
the parameter space of O stars covered by these studies based on non-LTE, 
line blanketed models with winds remains only partly sampled. 
This problem can be alleviated by the computation of 
a grid of models covering the whole range of parameters from which 
``theoretical calibrations'' can be derived.
However, as in principle the parameter space -- accounting e.g.\
for \mdot, the velocity law, microturbulence and others -- is rather large, assumptions
have to be made on the adopted input parameters or combinations thereof.
Finally, these models can also be tested by comparison to observational 
results from individual stars.
Given these advantages and limitations we have chosen to use both 
approaches to establish a new \teff\ scale and to examine their 
implications on the recalibration of the remaining stellar parameters.

In Sect.\ \ref{modelling} we describe the atmosphere models; we present 
the sample of observed stars from which we have derived estimated 
spectroscopic gravities and absolute visual magnitudes of dwarfs, giants 
and supergiants in Sect.\ \ref{obs_sample}; in Sect.\ 
\ref{results} we explain our method to derive the effective temperature scales; 
in Sect.\ \ref{other_param}, we show the results of the calibrations of 
other stellar parameters (bolometric correction, luminosity, radius, mass, ionising flux);  
in Sect.\ \ref{disc_Z} we discuss the effect of metallicity on the present results
 and the conclusions are gathered in Sect.\ 
\ref{conc}.

%%%%%%%%%%%%%%%%%%%%%%%%%%%%%%%%%%%%%%%%%%%%%%%%%%%%%%%%%%%%%%%%%%%%%%%%
%----------------------------------------------------------------------
%
%         Modelling 
%
%----------------------------------------------------------------------
\section{Modelling}
\label{modelling}

We have constructed a grid of models for O stars using the atmosphere 
code CMFGEN (Hillier \& Miller \cite{hm98}). We have tried to sample as 
well as possible the \logg--\teff\ plane in order to cover the whole 
range of spectral types and luminosity classes. In practice, we have 
relied on the grid of CoStar models of Schaerer \& de Koter 
(\cite{costar3}): models A2$\rightarrow$E2, A3$\rightarrow$E3 and
A4$\rightarrow$D4 have been recomputed 
with the same set of parameters as in the original grid. In addition, 
we have computed new models to refine this grid:  after adopting \teff\ 
and \logg, we have derived \lL\ from the 
evolutionary tracks at solar metallicity of Meynet et al.\ (1994),
which lead to $R$ and then $M$ through \logg; the 
terminal velocity (\vinf) was chosen to be 2.6 time the escape velocity 
(see Lamers et al.\ \cite{lamers95}, Kudritzki \& Puls \cite{kp00}) and 
the mass loss rate was 
computed according to the cooking recipe of Vink et al.\
(\cite{vink01}). A total of 38 models with $3.0 < \logg < 4.25$ and
$25000 < \teff\ < 48000$ K was finally computed.

The main characteristics of CMFGEN are largely described by Hillier 
\& Miller (\cite{hm98}). We give here a short overview:

- {\em Non-LTE:} all statistical equilibrium equations are solved individually 
to provide the detailed level populations. The radiative transfer 
equation is also solved and leads to the radiation field.

- {\em Line-blanketing:} line-blanketing is included directly by means of a 
super-level approach. In addition to H and He, C, N, O, Si, S and Fe 
level populations have been computed in our models. This amounts to 
$\sim$ 2000 levels and $\sim$ 800 super-levels. Solar abundances 
according to Grevesse \& Sauval (\cite{gs98}) have been adopted.
Numerous tests have been undertaken to confirm the accuracy of the 
super-level approach. While the strength of individual lines can be affected 
by using too few super-levels, or by poor super-level assignments, the method 
provides consistent non-LTE blanketed structures. Moreover, the super-level 
assignments are easily changed, and are often simply used to group levels of 
the same LS configuration. Note that the adopted super-level approach does not 
directly influence line blanketing: its influence is indirect, and arises 
through its effect on the non-LTE level populations obtained from the solution 
of the statistical equilibrium equations
%\textbf{Note that the use of super-levels is the only approximation of the 
%code as regards the treatment of line-blanketing. This approximation can easily 
%be dropped provided the computational resources are available (memory size). 
%Although 
%some subtle effects related to the use of super-levels may exist, we believe 
%that most of the line-blanketing and 
%blocking is reproduced by this approach. Notice also that for H and He, no 
%super-levels are used.}

- {\em Temperature:} the temperature structure is derived from the radiative 
equilibrium equation. 

\begin{figure}
\centerline{\psfig{file=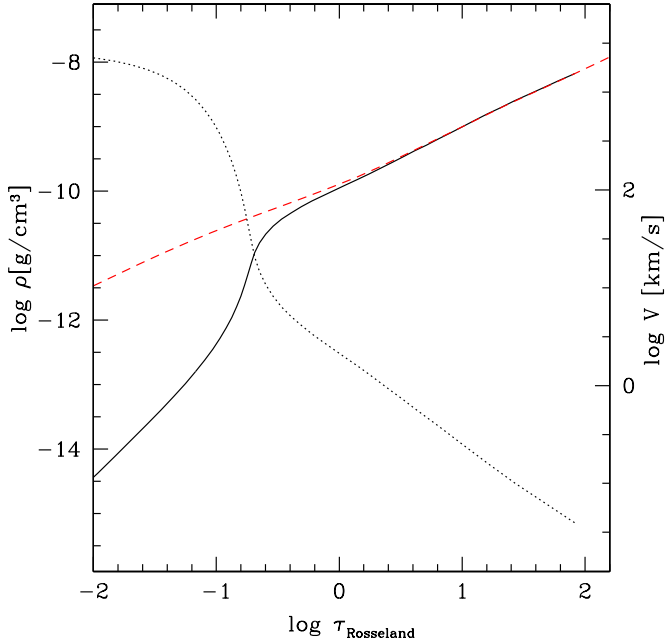,width=9cm}}
\caption{Typical hydrodynamical structure: density as a function of Rosseland 
optical depth for a model with \teff\ = 40000 K, \logg\ = 3.50 and \mdot\ = 
$1.4 10^{-5}$ M$_{\odot}$/yr. The red dashed 
line is the initial TLUSTY structure on which a $\beta$ velocity law is connected 
to give the final CMFGEN input structure (black solid line). The dotted line shows 
the CMFGEN velocity structure as a function of Rosseland optical depth. The TLUSTY 
and CMFGEN structures are similar in the very inner atmosphere and start to depart 
from each other around $\log \tau_{Rosseland} \sim 0$ which corresponds to a velocity 
of a few \kms. 
}
\label{comp_struc}
\end{figure}

- {\em Hydrodynamical structure:} CMFGEN does not compute self-consistently 
the hydrodynamical structure which has to be given as input and which 
is held fixed. We have adopted the velocity 
structure resulting from the smooth connection 
by smooth connection, 
we mean that both the velocity $v(r)$ and its derivative $dv/dr$ are imposed 
to be continuous at the connecting point.
of a pseudo-hydrostatic 
photospheric velocity structure with the $\beta$-law ($v =
v_{\infty}(1-\frac{R}{r})^{\beta}$) representing the wind (with a value 
of $\beta$ chosen to be 0.8).
The photospheric structure was taken from the TLUSTY grid of models 
OSTAR2002 (Lanz \& Hubeny \cite{ostar2002}).
Fig.\ \ref{comp_struc} shows the initial TLUSTY structure 
(dashed line) together with the CMFGEN input structure (solid line) in 
a typical case. 
One sees that the transition between the TLUSTY hydrostatic structure 
and the CMFGEN ``wind'' structure corresponds to a velocity of the 
order of a few \kms\ (or typically 10 to 30 $\%$ of the sound speed) which ensures 
that the important effects of velocity fields and spherical expansion are taken 
into account even in the photosphere (e.g. Kudritzki \cite{kud98}). Note also that 
the density at the connecting point scales with the mass loss rate, so that the 
adopted wind structure has most effects in supergiant models. 
When the parameters \teff--\logg\ did not match any of 
the TLUSTY models, we have simply linearly interpolated the photospheric 
velocity structure from the closest neighbours in the OSTAR2002 grid 
(see also Martins et al.\ \cite{wwgal}). For example for a model 
with temperature $35000 < \teff\ < 40000$ K, we estimated the density and 
optical depth from the TLUSTY models with \teff\ = 35000 and 40000 K according 
to $x = \frac{\teff -35000}{40000-35000}*(x40-x35)+x35$ where $xi$ is 
the density or optical depth at temperature $i$ in kK. This method, although 
approximate, was tested by Bouret et al.\ (\cite{jc03}) and was found to give the 
same structure as a true computation with TLUSTY.

- {\em Microturbulent velocity:} a microturbulent velocity can be included 
in the models. In the computation of the atmospheric structure (level 
populations, radiation field), we have chosen a value of 20 \kms\ while 
in the computation of the formal solution of the radiative transfer 
leading to the detailed emergent spectrum, we have adopted $v_{turb}$ 
= 5 \kms. The quite high value used in the atmospheric structure is 
required to avoid extremely long computation time. Test models have revealed 
that the \hei\ and \heii\ line profiles were basically unaffected by a decrease 
of \vturb\ from 20 \kms\ to 10 \kms\ in the atmosphere structure calculation 
(see Martins et al.\ \cite{n81}, Martins 
et al.\ \cite{wwgal}). Note also that microturbulent velocities of 10 to 20 
\kms\ are often found when modelling supergiants (e.g. hillier et al.\ 
\cite{hil03}).

%%%%%%%%%%%%%%%%%%%%%%%%%%%%%%%%%%%%%%%%%%%%%%%%%%%%%%%%%%%%%%%%%%%%%%%%
%----------------------------------------------------------------------
%
%        Observed sample
%
%----------------------------------------------------------------------
\section{Observed sample}
\label{obs_sample}

We have compiled the results from  
optical spectroscopic analysis of Galactic O stars using non-LTE spherically 
expanding models including line-blanketing. In practice, we have used the 
results of the studies of Herrero et al.\ (\cite{hpn02}) and Repolust et 
al.\ (\cite{repolust}) based on models computed with FASTWIND. We have also 
used our own results from the analysis of Galactic O dwarfs with CMFGEN 
(Martins et al.\ \cite{wwgal}) so that the number of O stars in this observed 
sample amounts to 45 objects. We have not included the results of Markova 
et al.\ (\cite{mark04}) since most of the stellar parameters where derived 
from calibrations and not from analysis with atmosphere models. 
We also excluded objects analysed by Bianchi \& Garcia 
(\cite{bg02}) and Garcia \& Bianchi (\cite{gb04}) since their studies are 
based on pure UV analysis and yield quite discrepant results to optical and 
combined UV-optical analysis for reasons we discuss now.

The above mentioned pure UV analysis (using WM-BASIC) 
relied on two main \teff\ indicators, \ov\ and \pv, lines 
which strongly depend on the wind properties (in particular clumping, see 
Crowther et al.\ \cite{paul02}, Bouret et al.\ \cite{jc05}). Moreover, 
WM-BASIC does not allow a treatment of Stark broadening which is crucial 
to correctly predict the width of gravity sensitive lines which are moreover 
essentially absent in the UV range. Note that the effects of clumping should 
not directly modify the present results, since we use photospheric lines formed well 
below the radius where clumping appears. In spite of this, we can not discard 
the fact that UV \teff\ diagnostics sometimes lead to lower values than optical 
lines, as is the case of the WM-BASIC studies of Bianchi \& Garcia 
(\cite{bg02}) and Garcia \& Bianchi (\cite{gb04}). However, other recent analysis 
show that consistent results with good fits can be obtained for 
\textit{photospheric} UV and optical lines (Bouret et al.\ \cite{jc03}, 
Hillier et al.\ \cite{hil03}, Martins et al.\ \cite{wwgal}).
Given this, and since our present approach 
is based on \teff - sensitive optical He lines, we do not include the
results of spectroscopic analysis based only on UV \textit{non-photospheric} lines
in our observational sample.
%which are less reliable for temperature diagnostisc (see Sect.\ \ref{theovsobs})
%and from which gravities are difficult to derive.   

%%%%%%%%%%%%%%%%%%%%%%%%%%%%%%%%%%%%%%%%%%%%%%%%%%%%%%%%%%%%%%%%%%%%%%%%
%----------------------------------------------------------------------
%
%         Gravities & Effective temperatures
%
%----------------------------------------------------------------------
\section{Gravities and effective temperatures}
\label{results}

As gravity is the main parameter determining the luminosity class 
a calibration of this quantity is needed to allow the calibration of
effective temperature for different LC.
Once established, such a \logg--ST relation for all LC can be used 
in the model atmosphere grid to produce new relations between effective 
temperature and spectral type, using gravity to
discriminate between luminosity classes.
These steps, as well as the determination of the observational \teff\
scale are now discussed.

All the resulting calibrated stellar parameters are summarised
in Tables \ref{tab_V} to \ref{tab_I_obs}.

%----------------------------------------------------------------------
\subsection{\logg --ST relation}
\label{loggST}

\begin{figure}
\centerline{\psfig{file=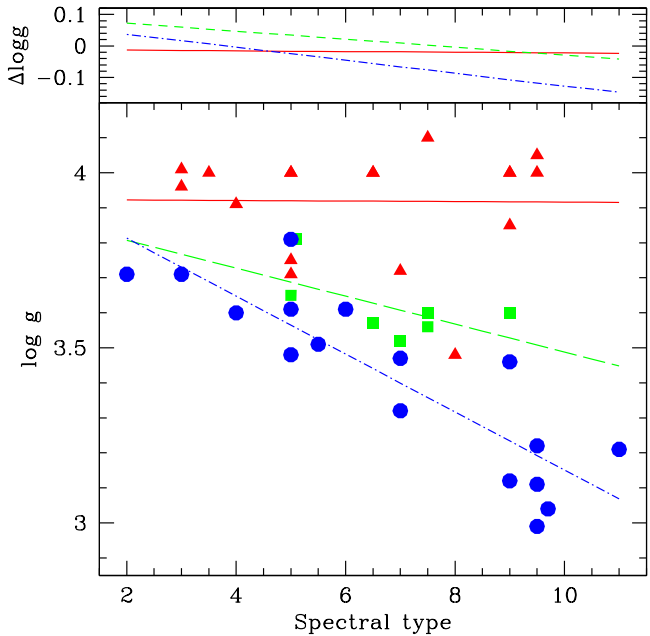,width=9cm}}
\caption{\textit{Lower panel}: gravity as a function of spectral type. 
Symbols corresponds to results 
derived from spectroscopic analysis of observed Galactic stars (sources: Repolust 
et al.\ \cite{repolust}, Herrero et al.\ \cite{hpn02} and Martins 
et al.\ \cite{wwgal}). Triangles are dwarfs, squares are giants and circles are 
supergiants. Lines are the least square fits of the observational data (solid: 
dwarfs; dashed: giants; dot-dashed: supergiants). \textit{Upper panel}: 
difference between the calibration of Vacca et al.\ (\cite{vacca}) and the present ones.
}
\label{logg_ST}
\end{figure}

The observational sample including recent spectroscopic analysis of individual 
stars has been used to derived an empirical 
calibration \logg--ST.
Note that the gravities used here are corrected for the effect of 
rotation, except for the dwarfs stars studied by Martins et al.\ (\cite{wwgal}). 
However in the latter study, the rotational velocities are not larger than 
130 \kms\ so that 
the difference between derived \logg\ and ``true'' \logg\ is negligible 
(see e.g. Repolust et al.\ \cite{repolust}).
 The results are shown in Fig.\ \ref{logg_ST} 
where dwarfs are represented by triangles, giants by squares and 
supergiants by circles. We note that there is a significant scatter 
in the empirical points leading to standard deviations of 0.15, 0.07 
and 0.12 dex respectively for dwarfs, giants and supergiants. The linear fits 
are to the observational results for dwarfs, giants and supergiants are:

\begin{equation}
\label{reg_logg}
\log g = \left\{ \begin{array}{l}
 3.924-0.001 \times \textrm{ST \hspace{1cm} (V)} \\
 3.887-0.040 \times \textrm{ST \hspace{1cm} (III)} \\
 3.979-0.083 \times \textrm{ST \hspace{1cm} (I)}
 \end{array} \right.
\end{equation}

\noindent where $ST$ is the spectral type.

Vacca et al.\ (\cite{vacca}) established a relation between spectroscopic 
gravities and spectral type from results of spectroscopic analysis through 
plane-parallel pure H He non-LTE models. We have proceeded similarly except that we 
have relied on atmosphere models including also winds and line-blanketing. Figure 
\ref{logg_ST} (upper panel) shows the differences between the two calibrations for 
various luminosity classes. We see that they are very close, the 
difference being less than 0.1 dex, which is the typical error generally 
quoted for the determination of spectroscopic gravities from the fit of 
Balmer lines. Herrero et al.\ (\cite{hpv00}) showed that the inclusion of winds in 
atmosphere models lead to a systematic increase of \logg\ by 0.1 to 0.25 dex. The 
present results show that on average gravities derived with plane parallel H He 
models are similar to those derived with line blanketed spherically extended models. 
Note that the results of Herrero et al.\ (\cite{hpv00}) are for supergiants with 
strong winds (\mdot\ of the order of $10^{-5..-6}$ M$_{\odot}$/yr) for which an 
effect of the wind on the \logg\ sensitive lines is not unexpected. For stars with 
such high density winds, the lowering of \teff\ imposed by the inclusion of 
line-blanketing may compensate for this wind effect so that, fortuitously, the 
relation \logg\ -- ST is not significantly changed.
%This may indicate that the effects of winds and line blanketing almost cancel. 
%This 
%statement should be tested further by analysis of individual stars with different 
%types of models.

%----------------------------------------------------------------------
\subsection{\teff scales}
\label{teffscales}

As already discussed, 
the derivation of new effective temperature scales was made following two 
different methods:
First, the relation between spectral type and \teff\ was derived directly 
from our grid of models, using the above \logg-- ST relation to 
discriminate between luminosity classes (``theoretical scale''). Second, 
we used the sample of 
stars analysed with quantitative spectroscopy to estimate the mean 
relation \teff--ST (``observational scale''). 
The resulting calibrated stellar parameters using the theoretical
\teff\ scale are summarised
in Tables \ref{tab_V} to \ref{tab_I}, those based on the 
observational \teff\ scale in Tables  \ref{tab_V_obs} to \ref{tab_I_obs}.

\subsubsection{\teff\ scale from models (``Theoretical scale'')}
\label{teffscaletheo}

Here, we show how we derived new \teff\ scales from 
our grid of models. 
The advantage of using models is that we can sample 
the whole spectral and luminosity range. However, the results depends 
on the grid of model and the parameters adopted to compute these models.

\begin{figure}
\centerline{\psfig{file=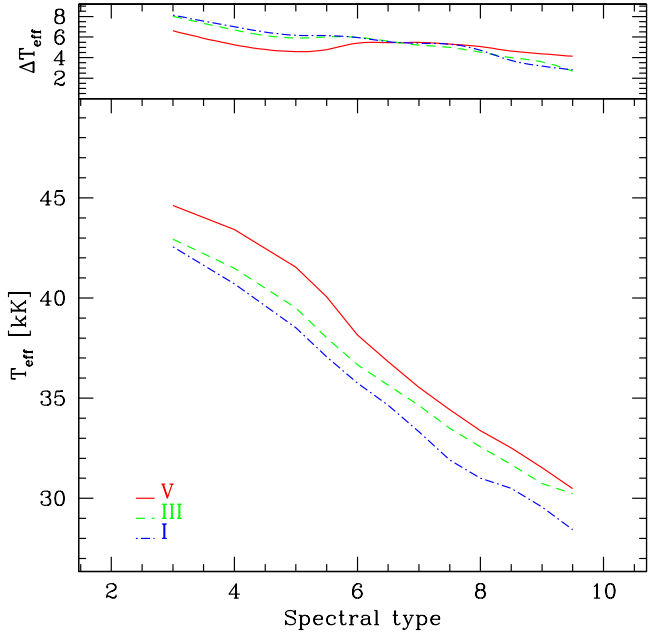,width=9cm}}
\caption{\textit{Lower panel}: ``theoretical'' effective temperature scale for 
dwarfs (solid line), 
giants (dashed line) and supergiants (dot-dashed line) as derived 
from our model grid. \textit{Upper panel}: difference between the 
\teff\ scales of Vacca et al.\ 
(\cite{vacca}) and the present results.
}
\label{Teff_ST}
\end{figure}

The basic idea is to build a function \teff(ST,\logg) from which 
temperatures from dwarfs, giants and supergiants --- characterised by a 
given relation \logg--ST --- were derived. For this, we have
made 2D interpolations using the NAG routines e01sef and e01sff. 
The basic principle of such routines is to construct a surface 
\teff(\logg,ST) regularly sampled in \logg--\teff from the grid model 
points (not regularly sampled in the space parameter) and then to 
determine the effective temperatures from this surface for a given 
\logg--ST relation. The construction of the surface involves 
two parameters representing basically the number of neighbouring input 
points from which we want the interpolation to be made for a given 
(\logg,\teff) of the regular grid. We have tested the effects of a 
change of the values of these parameters on our results and have found 
that for a reasonable number of neighbouring points ($>$ 15) there was 
essentially no modification. Note that we have not used directly the 
spectral type as a parameter, 
but we rely on the ratio of the equivalent widths  
$\log (EW(HeI \lambda 4471)/EW(HeII \lambda 4542)) \equiv \log W'$), 
which is directly related to ST through the classification scheme 
of Mathys (\cite{mathys88}).
According to this scheme, a given spectral type corresponds to a 
range of values for $W'$ so that all stars having $W'$ in this range 
will have the same spectral type, introducing thus a dispersion in 
any calibration as a function of ST. 
This interpolation process leads 
to the \teff--ST relations displayed in Fig.\ \ref{Teff_ST}. The 1 
$\sigma$ dispersion on the \logg--ST relation translates to an error of 
$\pm 500 K$ for the late spectral types and up to $\pm 1000 K$ for 
early spectral types, with little differences between luminosity classes. 
Note that between spectral types O9.5 and O9.7 there is only 0.2 unit 
in terms of spectral type, but the difference in terms of $\log W'$ 
is 0.3 dex, which is as large 
as between spectral types O7.5 and O9. Hence, we have excluded 
the O9.7 spectral type in our calibrations since the interpolations 
(made on $\log W'$ and not on spectral type) lead to artificially 
low effective temperatures. 

The differences between our temperature scales and the calibrations
of Vacca et al.\ (\cite{vacca}) are plotted in the upper panel of
Fig.\ \ref{Teff_ST}. Typically they reach from 2000 and 8000 K, 
the largest differences being found for supergiants.

We note also that our effective temperatures for 
the latest spectral types (O9.5) agree reasonably well with the effective 
temperatures 
of the earliest B stars. Indeed, Schmidt-Kaler (\cite{sk82}) predict 
\teff\ = 30000 K (29000, 26000) for dwarfs (giants, supergiants) of 
spectral type B0, while we derive \teff\ = 30500 K (30200, 28400) for 
O9.5 dwarfs (giants, supergiants).

\subsubsection{``Observational'' \teff\ scale from a sample of Galactic O stars}
\label{teffscaleempi}

\begin{figure}
\centerline{\psfig{file=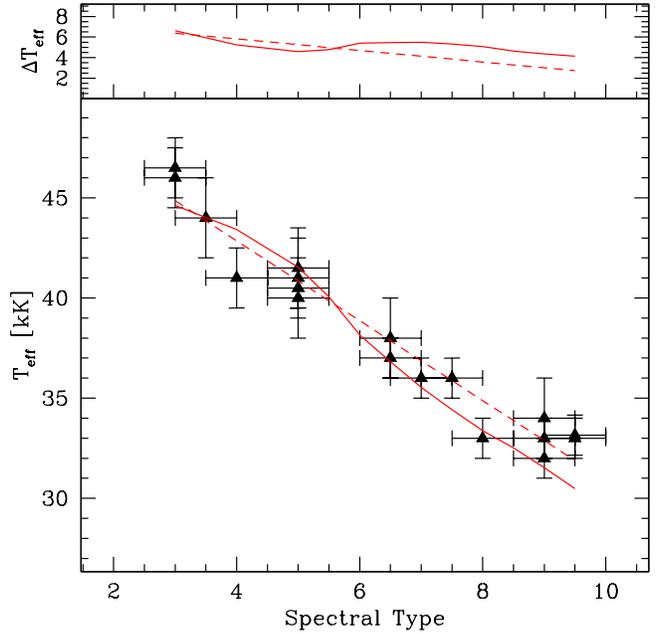,width=9cm}}
\caption{Effective temperature scale for dwarfs as derived from our model 
grid (``theoretical'' scale, solid 
line) compared to ``observational'' results based on non-LTE line blanketed models 
including winds (Herrero et al.\ \cite{hpn02}, Repolust et al.\ \cite{repolust}
and Martins et al.\ \cite{wwgal}). The dashed line is the result of linear 
regression to the data points.
The typical error bars for the observational 
results are $\pm 0.5$ unit for spectral types and are taken from detailed 
spectroscopic analysis for \teff.
The upper panel shows the difference between the relation of Vacca et al.\ 
(\cite{vacca}) and our \teff\ scales (solid line: relation from model grid; 
dashed line: relation from spectroscopic analysis). }
\label{Teff_ST_V}
\end{figure}

\begin{figure}
\centerline{\psfig{file=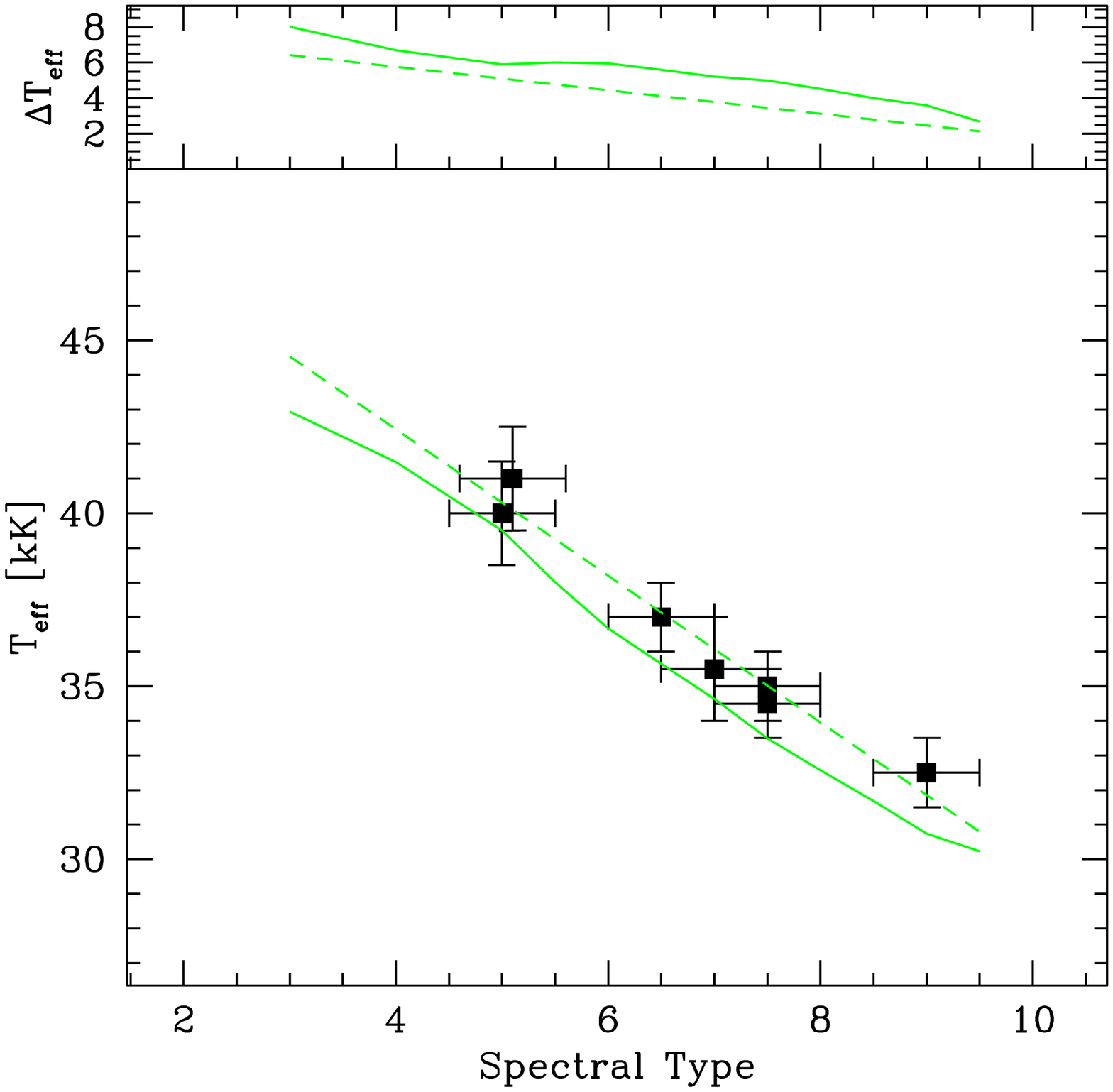,width=9cm}}
\caption{Same as Fig.\ \ref{Teff_ST_V} for giants.}
\label{Teff_ST_III}
\end{figure}

\begin{figure}
\centerline{\psfig{file=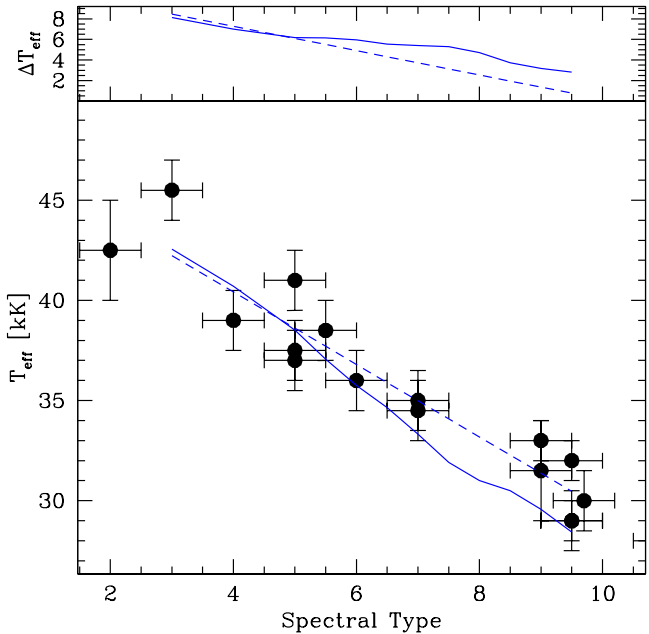,width=9cm}}
\caption{Same as Fig.\ \ref{Teff_ST_V} for supergiants.}
\label{Teff_ST_I}
\end{figure}

Instead of relying on a grid of models, one can simply derive mean 
\teff\ scales from the results of detailed spectroscopic analysis 
of O stars (see also Repolust et al.\ \cite{repolust}). 
The advantage is that the parameters of the best models 
are hopefully the true stellar and wind parameters of the stars analysed. 
But the problem is that the number of objects is limited and that all 
spectral types are not available (this is especially true for giants). 
In Figs.\ \ref{Teff_ST_V}, \ref{Teff_ST_III} and \ref{Teff_ST_I}, we show 
the results of a linear regression to the data points (dashed line) in 
comparison to the \teff\ scale derived from the grid of models (solid line).

The linear fits to the observed data are:

\begin{equation}
\label{reg_teff}
\teff\ = \left\{ \begin{array}{l}
 50838 - 1995 \times \textrm{ST \hspace{1cm} (V)} \\
 50882 - 2115 \times \textrm{ST \hspace{1cm} (III)} \\
 47666 - 1881 \times \textrm{ST \hspace{1cm} (I)}
 \end{array} \right.
\end{equation}

\noindent and the dispersions are 1021, 529 and 1446 K respectively.

The differences with the relations of Vacca et al.\ (\cite{vacca}) are 
also shown in the upper panels of the figures and are slightly smaller 
(2000 K) than for the theoretical scale, mainly for late spectral types.

\subsubsection{Theoretical vs observational \teff\ scale}
\label{theovsobs}

At this point, it is worth discussing the differences between the 
``theoretical'' 
and the ``observational'' relation.
First of all, given the uncertainties and the natural 
dispersion of the observational results ($\sim$ 2000-3000 K) the agreement 
between our theoretical and observational scales is good for dwarfs and 
supergiants, although our effective temperatures seem to be slightly 
underestimated (by $\sim$ 1000-2000 K) for late spectral types in our models.
For giants, our theoretical \teff\ scale seems to be a little too cool (by $\sim$ 1500 K). 
Note, however, that the number of giants studied so far is relatively small (7). 

The following assumptions could be at the origin of the lower effective 
temperatures in our models compared to the results from the analysis of 
individual O stars:
{\em 1)} incorrect metallicity/abundances, 
{\em 2)} incorrect atmosphere and wind parameters (microturbulent velocities, 
	mass loss rates, slope of the wind velocity law), or 
{\em 3)} incorrect gravities.
We now discuss these one by one.

{\bf 1) Metallicity/abundances:} 
First, it is well established that line-blanketing leads 
to lower \teff\ so that a too strong decrease could be due to a too high 
metal content. However, all stars used for the comparison are Galactic 
stars which should have solar composition. Of course there is a metallicity 
gradient in the Galaxy, but the metal content remains within approximately
$\pm$ 2 times 
the solar content (e.g. Giveon et al. \cite{giveon}). Since 
we have shown that increasing the effective 
temperature by 1000 to 2000 K required a decrease of the metal content 
by a factor of 8 (paper I and Martins \cite{thesis}), it is not likely 
that the difference between our new \teff\ scales and observational 
results can be explained by different metal contents.
 
{\bf 2) Atmosphere and wind parameters:} 
Another possibility is that some atmospheric and wind parameters used for our 
modelling of O stars atmosphere are not exactly the same as the parameters 
of real O stars. In paper I, we have tested a number of such parameters 
(microturbulent velocity, mass loss rates, slope of the velocity field 
in the wind -$\beta$ parameter-) and shown that increasing \vturb, 
decreasing $\beta$ or decreasing \mdot\ could lead to later spectral 
types for a given \teff\ (or equivalently to higher \teff\ for a given 
spectral type). 
Since $\beta$ = 0.8 is already small in our models, it 
is not likely that this parameter can explain the difference. However, 
mass loss can explain the discrepancy since 
most of the late type observed dwarfs in Fig\ \ref{Teff_ST_V} 
are from Martins et al.\ (\cite{wwgal}) and have \mdot\ much lower 
than the predictions of the Vink et al.\ (\cite{vink01}) cooking recipe 
used to assign a mass loss rates in the present models. Indeed, Martins 
et al.\ (\cite{wwgal}) found that dwarfs with luminosities below 
$\sim 10^{5.2}$ L$_{\odot}$ have lower mass rates than the results of 
hydrodynamical simulations. But Fig.\ \ref{Teff_ST_V} shows that the 
discrepancy between our derived effective temperatures and the observed 
ones starts for stars with spectral types later than O6, and it turns 
out that according to our new calibrations an O6V star has a luminosity 
of $\sim$ 10$^{5.3}$ L$_{\odot}$. We have also shown in paper I that 
increasing \mdot\ by a factor 2 lead to a decrease of $\log W'$ by $\sim$ 0.1 dex.
Hence, it is likely that mass loss 
is at the origin of the underestimation of \teff\ in our models of 
late type dwarfs compared to results of spectroscopic analysis. 
Concerning supergiants, the situation is a little more complicated 
since according to Table \ref{tab_I} all supergiants have luminosities 
higher than 10$^{5.2}$ L$_{\odot}$ and should therefore not have reduced 
winds. Hence, the reason for the lower \teff\ in our models is probably 
not rooted in the mass loss rates. 

Mokiem et al.\ (\cite{mokiem}) have also
shown, from the computation of theoretical spectra, that an increase 
of microturbulent 
velocity from 5 to 20 \kms\ translates to a shift towards later spectral 
types by nearly half a sub-type since both He~{\sc i} and He~{\sc ii} 
lines are stronger but the former are more affected (cf.\ paper I),
although the effect 
is usually more important on singlet than on triplet  He~{\sc i} lines 
(see also Smith \& 
Howarth \cite{sh98}, Villamariz \& Herrero \cite{vh00})). 
This can partly explain the observed 
behaviour for supergiants for which microturbulent velocities of 20 \kms\ 
are possible. Test computation in our models reveal that increasing \vturb\ 
by such an amount corresponds to a shift in $\log W'$ by $\sim$ 0.15, 
which is of the order of the width of the range of $W'$ values within a 
spectral type. 
Note also that in our grid of models, we have used \vturb\ = 20 \kms\ to 
compute the atmospheric structure. This may be a little to high for dwarfs,
but as already discussed (Sect.\ \ref{modelling}) test models reveal 
that this does not 
lead to an overestimation of the line-blanketing effect (only some particular 
lines are modified). 
%with the consequence of an increased effect of line-blanketing. However, 
%this may also compensate for the lack of additional species (Ni, Al, Ca...) 
%which can contribute to line-blanketing (although moderately, see paper I). 
%If the effect of a too large microturbulent velocity is larger than the effect 
%of a lack of additional metals, then we should expect a slightly too cool 
%\teff\ - scale. However, this uncertainty should remain within one spectral 
%subtype. 

{\bf 3) Gravity:} 
Another possibility to explain the different \teff\ - scales
is that gravities from spectroscopic analysis may be underestimated. 
This would lead to lower gravities for a given luminosity class, and 
consequently to lower effective temperatures.   
However, if we recompute the \teff\ 
scale with gravities increased by the standard deviations obtained 
from least square fits of observed data, the corresponding increase 
in \teff\ we find ($\sim$ 1000 K, see Sect.\ \ref{teffscales}) is not
sufficient to explain the difference with 
observed effective temperatures. This is especially true for giants for 
which the theoretical effective temperatures are systematically 
underestimated by $\sim$ 1500 K. Note however that the small number 
of giants studied so far may hide the natural spread in the \teff\ -- 
spectral type relation we see for dwarfs and supergiants. This may 
thus artificially increase the difference with our theoretical results.

We also want to mention that we have not included the results of 
Bianchi \& Garcia (\cite{bg02}) and Garcia \& Bianchi (\cite{gb04}) 
in our comparisons since they are based on UV analysis, whereas both 
the present modelling and the observational results of Repolust et al.\ 
(\cite{repolust}), Herrero et al.\ (\cite{hpn02}) and Martins et al.\ 
(\cite{wwgal}) rely on optical diagnostics. Bianchi \& Garcia (\cite{bg02}) 
and Garcia \& Bianchi (\cite{gb04}) found effective temperature much 
lower (by several thousands K) compared to any optical study. This may 
be partly due to the dependence of the UV diagnostics on several parameters 
other than \teff, especially clumping which may significantly alter the 
ionisation structure in the wind \textbf{(but not the photosphere)} and 
thus the strength of UV lines 
and which is not taken into account in their analysis. 

\begin{figure}
\centerline{\psfig{file=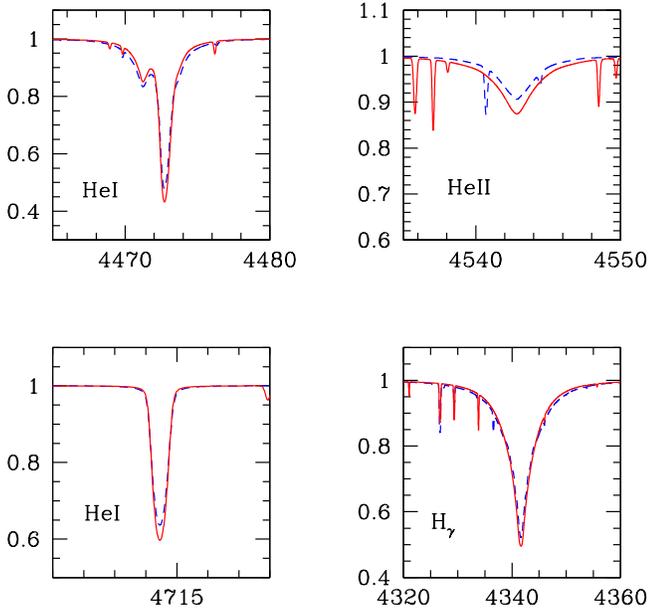,width=9cm}}
\caption{Comparison between H and He line profiles of the same model (A2, 
see paper I) computed with the Isa-Wind photospheric structure (dashed line) 
and the TLUSTY structure (solid line). See text for discussion.}
\label{comp_HHe_lines}
\end{figure}

We note that the 
present \teff\ scale for O dwarfs is slightly 
cooler ($\sim$ 500 K) than that of paper I. This is explained first 
by the lower 
gravities used here for dwarfs (values of 4.00-4.05 were adopted for 
$\log g$ in paper I) and second by the different pseudo-hydrostatic 
structure used: ISA-Wind structures were computed in paper I while TLUSTY 
structures are used here. The difference is illustrated in Fig.\ 
\ref{comp_HHe_lines} where we see the effect of the photospheric structure 
on the He classification lines in a model with \teff\ = 33300 K. TLUSTY 
structures, which are more realistic due to a better treatment of 
line-blanketing, have higher temperature and thus a higher thermal pressure 
which implies a higher density (through the hydrostatic equilibrium equation). 
This leads to slightly higher ionisations and then to earlier 
spectral types. 
The present results should represent an improvement over paper I.

The general conclusion of the above discussion is that our theoretical \teff\ 
scale should be taken as indicative and have an uncertainty of 
$\pm$ 1000 to 2000 K due to both a natural dispersion and 
uncertainties inherent to our global method. Hence in the following, we 
use both these theoretical \teff\ scales AND the relation derived from 
detailed modelling of individual stars (``observational'' relations) to 
calibrate the other stellar parameters.

%%%%%%%%%%%%%%%%%%%%%%%%%%%%%%%%%%%%%%%%%%%%%%%%%%%%%%%%%%%%%%%%%%%%%%%%
%----------------------------------------------------------------------
%
%         Other stellar parameters
%
%----------------------------------------------------------------------
\section{Other stellar parameters}
\label{other_param}

We now derive calibrations for the remaining principal stellar parameters
based on the two temperature scales determined above. 
The results are summarised in Tables \ref{tab_V} to \ref{tab_I_obs}.

%----------------------------------------------------------------------
\subsection{Magnitudes, bolometric corrections, and luminosities}
\label{BC}

The calibration of luminosity requires two ingredients: first the 
absolute visual magnitude and second the bolometric corrections. To 
calibrate \mv\ as a function of spectral type, we can either rely 
entirely on models and proceed exactly as for the effective temperature, 
i.e. adopting the \logg--ST relations, or we can derive empirical 
calibrations of \mv\ from results of analysis of individual stars, 
as we have done for \logg. Here, we have chosen the latter method. 
The advantage is that the derived \mv--ST calibration depends 
directly on observations and relies less on atmosphere models. It 
may however suffer from the often poorly know distances of the analysed 
stars, rendering more uncertain the absolute magnitudes. The former 
method (\logg--ST + \mv\ from models) relies on better \mv\ but 
suffers from the uncertainty in the \logg--ST calibration. As the 
final goal is to produce calibrations as close as possible to 
observations, we do not retain this method. 

\begin{figure}
\centerline{\psfig{file=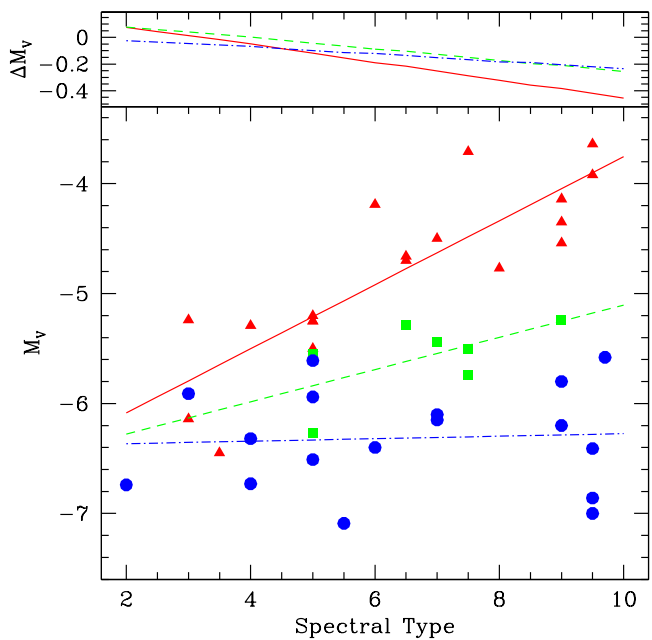,width=9cm}}
\caption{Same as Fig.\ \ref{logg_ST} for absolute magnitudes. 
The standard deviation from the mean relation is 0.4 mag for dwarfs, 
0.26 mag for giants and 0.45 mag for supergiants.}
\label{Mv_ST}
\end{figure}

The result of the empirical calibration \mv--ST is shown in Fig.\ 
\ref{Mv_ST} where the symbols and observational data are the same as 
in Fig.\ \ref{logg_ST}. The standard deviation is 0.40 for dwarfs, 
0.26 for giants and 0.45 for supergiants. The linear fits to the 
observational data are:

\begin{equation}
\label{reg_Mv}
M_{V}  = \left\{ \begin{array}{l}
 0.291 \times \textrm{ST - 6.667 \hspace{1cm} (V)} \\
 0.147 \times \textrm{ST - 6.570 \hspace{1cm} (III)} \\
 0.011 \times \textrm{ST - 6.388 \hspace{1cm} (I)}
\end{array} \right.
\end{equation}

\noindent where $ST$ is the spectral type. One may argue that 
due to the fact that massive stars evolve almost horizontally in the HR 
diagram after the main sequence, there could be a wide range of luminosities 
(and \mv) for a given \teff\ (or equivalently ST) for supergiants so that 
a calibration \mv\ -- ST for luminosity class I stars may be questionable. 
However, the observed dispersion observed in the present study and in the 
Vacca et al.\ (\cite{vacca}) study is not significantly different between 
dwarfs, giants and supergiants. Also, the observed samples show a clear 
separation in \mv\ between the different luminosity classes which is even 
more highlighted by the fact that we consider only 3 luminosity classes 
(and not 5 as usual). Hence, we think that a calibration \mv\ -- ST even 
for supergiants makes sense.
%It is worth noting 
%here that the relation for supergiants is only indicative since massive 
%in a post main sequence phase evolve almost horizontally in the HR 
%diagram (see Fig.\ \ref{HR_diag}) so that for a given \teff\ (or spectral 
%type) there could a 
%range of values for $L$ and \mv. This partly explains the large 
%dispersion of the \mv--ST relation in spite of the relatively large 
%number of supergiants studied so far.}

Absolute visual magnitudes have also been derived from observational results 
by a number of authors. 
The upper panel of Fig.\ \ref{Mv_ST} shows the difference between the 
calibrations of Vacca et al.\ (\cite{vacca}) and the present ones.
They do not differ by more than 0.4 dex which is also the typical 
dispersion of our relations. Hence, we can not quantitatively compare 
both types of relations: agreement or small differences ($<$ 0.4 dex) can not be 
distinguished.

Recently, Schr$\ddot{\rm o}$der et al.\ (\cite{schroder04}) have tested the 
\mv--ST relations 
of Schmidt-Kaler (\cite{sk82}), Howarth \& Prinja (\cite{hp89}) and Vacca et 
al.\ (\cite{vacca}) by comparing the parallaxes derived from such relations 
to the parallaxes measured by Hipparcos for a number of Galactic O stars. They 
concluded that the three calibrations are consistent with the Hipparcos 
measurement. As our relations are not strongly different from the relations 
of Vacca et al.\ (\cite{vacca})  
tested by Schr$\ddot{\rm o}$der et al.\ (\cite{schroder04}), they are certainly 
reasonable.

\begin{figure}
\centerline{\psfig{file=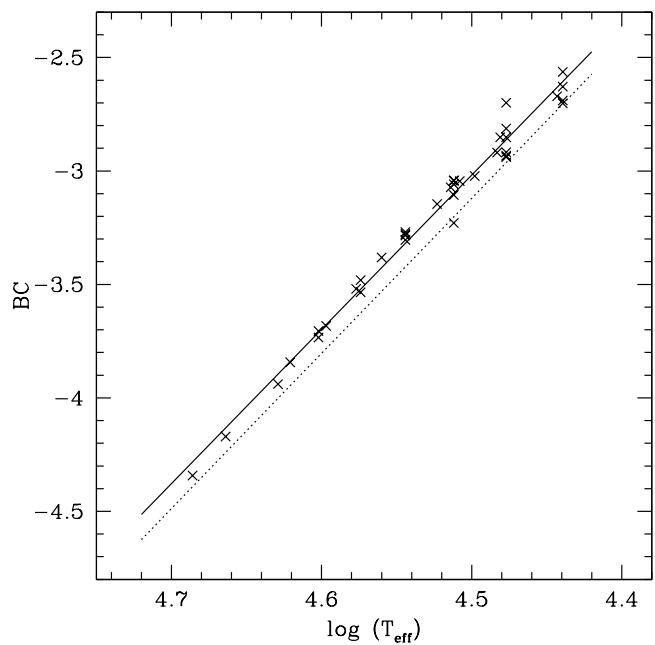,width=9cm}}
\caption{Bolometric correction as a function of effective temperature
from our set of CMFGEN models. The solid line is the linear regression 
(Eq.\ \ref{eq_BC_teff}) and the dotted line is the relation of Vacca et 
al.\ (\cite{vacca}) and is shifted by $\sim$ 0.1 mag compared to our relation.}
\label{BC_Teff}
\end{figure}

To obtain the luminosities from absolute magnitudes, one needs to 
know the bolometric correction. Vacca et al.\ (\cite{vacca}) have shown 
that bolometric corrections were essentially independent on the 
luminosity class and depended mainly on \teff. We have computed the 
BC of our models from luminosities and synthetic \mv, yielding
the following linear dependence on \teff:

\begin{equation}
BC = 27.58 - 6.80 \times \log \teff
\label{eq_BC_teff}
\end{equation}

\noindent with a standard deviation of 0.05.
Our relation is shifted by 0.1 magnitude compared to that of 
Vacca et al.\ (\cite{vacca}) 
as seen on Fig.\ \ref{BC_Teff}. This is explained by the effect of 
line-blanketing which redistributes the flux blocked at short wavelengths 
to longer wavelengths, including the optical. This translates to an 
increase of the V magnitude by $\sim$ 0.1 mag, and consequently to a reduction 
of the bolometric correction by the same amount (for a model of given 
luminosity).

%This similarity may be surprising since line-blanketing redistributes 
%flux from short to long wavelength, including optical wavelength. However, 
%in the latter range, the flux increase does not exceed 10 \%, which
%implies a reduction of the bolometric correction by less than 0.1 magnitude. 
%This is consistent with the dispersion of our relation BC--\teff. 

\begin{figure}
\centerline{\psfig{file=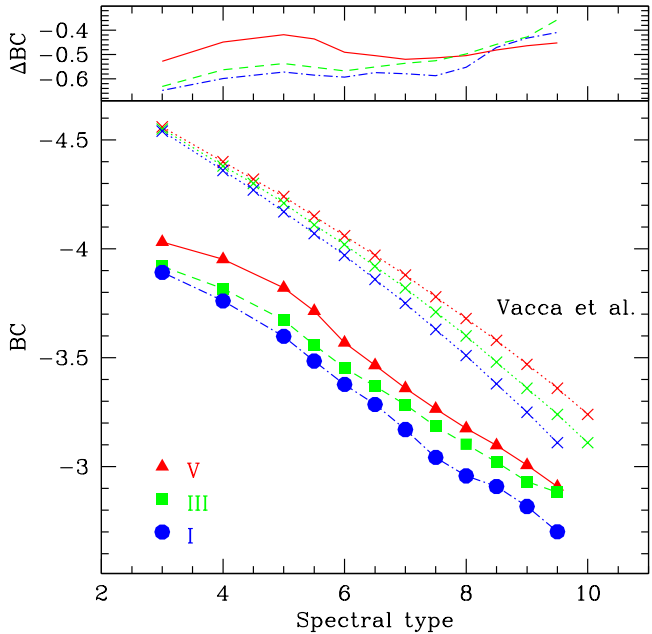,width=9cm}}
\caption{Bolometric correction as a function of spectral
type from our set of CMFGEN models (solid lines) and from Vacca et al.\ 
(\cite{vacca}, dotted lines). BCs are usually reduced by $\sim$ 0.2 to 0.5 
magnitudes for a given spectral type, the reduction being the largest 
for early supergiants.}
\label{BC_ST}
\end{figure}

\begin{figure}
\centerline{\psfig{file=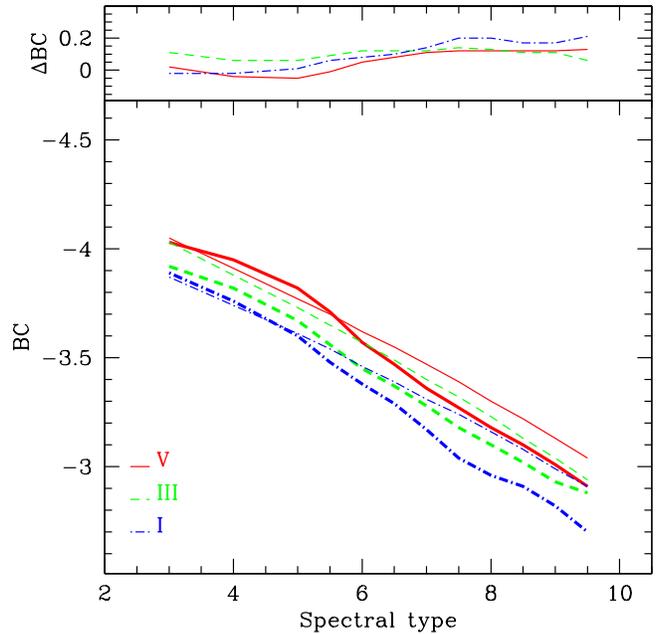,width=9cm}}
\caption{Same as Fig.\ \ref{BC_ST} but with the bolometric corrections 
derived using the ``theoretical'' (bold lines) and ``observational'' 
(thin lines) \teff\ scales. The upper panel shows the differences 
between ``theoretical'' and ``observational'' BCs.
}
\label{BC_ST_diff}
\end{figure}

Figure \ref{BC_ST} shows our calibration of bolometric corrections as 
a function of spectral type obtained with our theoretical 
\teff\ scale together with the relations of Vacca et 
al.\ (\cite{vacca}). The difference between both types of relations 
is displayed in the upper panel. The reduction of BC goes from $\sim$ 0.4 to 
$\sim$ 0.5 magnitude for dwarfs, from  $\sim$ 0.35 to $\sim$ 0.6 mag for giants 
and from $\sim$ 0.4 to $\sim$ 0.65 mag for supergiants.
Figure \ref{BC_ST_diff} shows the bolometric corrections estimated with both 
the ``theoretical'' and ``observational'' \teff\ scales. Both types of BCs 
differ by less than 0.2 magnitudes (which is lower than the 
difference with the results of Vacca et al.\ \cite{vacca}), except for the 
latest supergiants. The lower BCs at late spectral types for the theoretical 
relation simply comes from the lower effective temperatures given by the 
theoretical \teff\ scales.

\begin{figure}
\centerline{\psfig{file=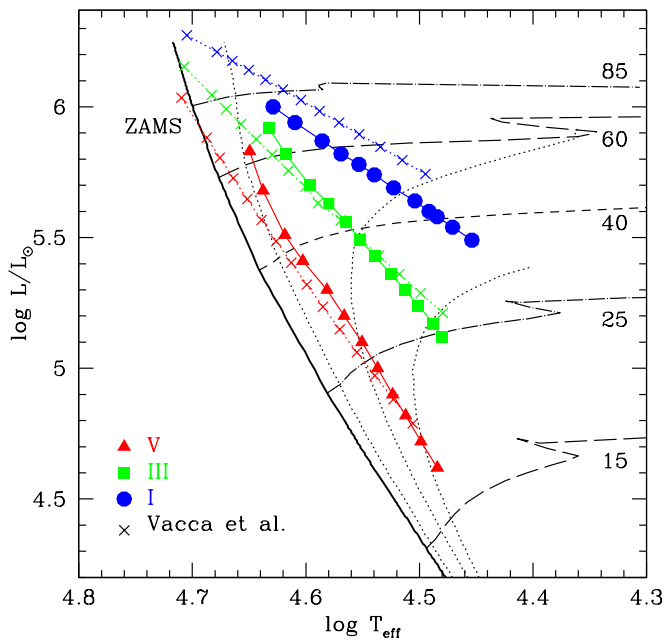,width=9cm}}
\caption{HR diagram showing the typical position of dwarfs (triangles), giants 
(squares) and supergiants (circles) from the present ``theoretical'' calibration of \teff\ and 
\lL. Isochrones for 0, 1, 3 and 5 Myrs are shown together with evolutionary tracks 
for different masses (taken from Meynet et al.\ \cite{meynet94}). 
The crosses 
and dotted lines are the positions of dwarfs, giants and supergiants following 
the Vacca et al.\ (\cite{vacca}) calibrations.}
\label{HR_diag}
\end{figure}

With the absolute magnitudes and bolometric corrections, 
the luminosities can be estimated according to 
\begin{equation}
\log \frac{L}{L_{\odot}} = -0.4 \times (\mv\ + BC - M^{bol}_{\odot})
\label{eq_lL}
\end{equation}

\noindent where $M^{bol}_{\odot}$ is the solar bolometric 
luminosity taken to be 4.75 
(Allen \cite{allen}). Again, both theoretical and observational \teff\
scales can be used. 
To estimate the error on the luminosities, we have recomputed them adding $\pm 
\sigma$ on \logg\ and \mv\ in the \logg--ST and \mv--ST  relations 
\textit{for a given \teff\ scale}. This means that these errors 
do not take into account the uncertainty on the effective temperatures. This 
results in errors of $\pm 0.15, 0.10$ and $0.17$ in \lL\ for dwarfs, giants 
and supergiants respectively.     
Figure \ref{HR_diag} shows the resulting average positions of dwarfs, 
giants and supergiants in the HR diagram. 
For a given spectral type, the luminosities are lower by 0.25-0.35 for 
dwarfs, by $\sim$ 0.25 dex for giants and by 0.25 to 0.35 dex for supergiants
compared to the values presented by Vacca et 
al.\ (\cite{vacca}). 
Given the small differences in BC for the two \teff\ scales it is clear
that $L$ depends little on this choice. Quantitatively, 
using the theoretical or observational \teff\ scales 
results in modifications of the luminosity by less than 0.1 dex for all LC.

A comment concerning the off-set of the dwarfs 
position compared to the Zero Age Main Sequence seen in Fig.\ \ref{HR_diag} is 
worthwhile. As we are determining the \textit{average} 
properties of O stars, it is not surprising that the O dwarfs relation 
does not fall on the Zero Age Main Sequence (ZAMS). However, a precise 
explanation of the offset, indicating \textit{average ages} of $\sim$ 1 to 
5 Myr for early to late O dwarfs, is still lacking.
A possible explanation could be that very young stars still embedded in their 
parental cloud may not be observable. As later types have weaker winds less 
efficient to blow this cloud, this could possibly even explain why later O 
dwarfs have older average ages. However, although few, some O dwarfs are found 
very close to the ZAMS (Rauw \cite{gregor}, Martins et al.\ \cite{wwgal}). 
Systematic studies of larger samples of the youngest O stars will be needed to 
settle this question.
Finally, the existence of a theoretical ZAMS is not fully established 
(Bernasconi \& Maeder \cite{bm96}) and the precise shape/location of the ZAMS 
and isochrones is also dependent on parameters like the rotation rate (Meynet 
\& Maeder \cite{mm00}).

%----------------------------------------------------------------------
\subsection{Mass and radius}
\label{M_R}

With the effective temperature scale and the luminosities as a function 
of spectral type, it is easy to derive the relation radius -- spectral type from

\begin{equation}
R = \sqrt{\frac{L}{4 \pi \sigma_{R} \teff^{4}}},
\label{eq_R}
\end{equation}

\noindent where $\sigma_{R}$ is the Stefan-Boltzmann constant. The standard
deviation on $R$ is 10 to 20 \%. 

Gravity can then be used to derive the spectroscopic mass according to

\begin{equation}
M = \frac{g R^{2}}{G},
\label{eq_M}
\end{equation}

\noindent with $G$ the gravitational constant. Masses must be taken
with care since their uncertainty is as high as 35 to 50 \%. 

For all results see Tables \ref{tab_V} to \ref{tab_I_obs}.

%----------------------------------------------------------------------
\subsection{Ionising fluxes}
\label{ion_flux}

\subsubsection{Calibrations}
\label{ion_flux_calib}

An important quantity related to massive stars is the number of ionising 
photons they emit due to their high effective temperature and luminosity. 
Ionising fluxes are defined as follows

\begin{equation}
q_{i} = \int^{\lambda_{i}}_{0} \frac{\pi \lambda F_{\lambda}}{h c} d \lambda
\label{eq_ion_flux}
\end{equation}

\noindent where $F_{\lambda}$ is the flux expressed in erg/s/cm$^{2}$/\AA\ and 
$\lambda_{i}$ is the limiting wavelength below which we estimate the 
ionising flux. In practice, we usually define $q_{0}$, $q_{1}$ and $q_{2}$ 
as the fluxes able to ionise Hydrogen ($\lambda_{0}$ = 912 \AA), He~{\sc i} 
($\lambda_{1}$ = 504 \AA) and He~{\sc ii} ($\lambda_{2}$ = 228 \AA). The 
ionising fluxes are highly sensitive to line-blanketing effects since metals 
have most of their bound-free transitions in the Lyman continuum. Hence, the 
presence of such bound-free opacities in metals modifies the spectral energy 
distribution (SED) and consequently the ionising fluxes: flux at high 
energy is blocked and must be re-emitted at longer wavelength. Moreover, the 
numerous metallic lines in this range create a veil which can also significantly 
alter the shape of the SED. 

Several studies have already provided new ionising fluxes as a function of 
effective temperature (Smith et al.\ \cite{smith}, Mokiem et al.\ \cite{mokiem}).
An in depth study of ionising fluxes as a function of spectral type and 
metallicity (and thus line-blanketing) was carried out by Kudritzki (\cite{kud02}) 
who found that the number of Lyman continuum photons was not a strong function of 
metallicity. 
%They have revealed that for a given \teff, the number of Lyman continuum photons 
%is usually close to those of previous models not including line-blanketing. 
%(\textbf{Abbott \& Hummer \cite{ah85}, Kudritzki \cite{kud02},}  Mokiem et al.\ \cite{mokiem}).
This may be surprising at first glance since, as we have already mentioned in Sect.\ \ref{BC}, 
flux redistribution takes place at all wavelengths, including $\lambda > 912$ 
\AA. However, energy flux and photon flux are not equivalent. It is possible for 
the flux in the Lyman continuum to go down while the photon flux stays almost 
constant. In this case, most of the photons reside close to the Lyman edge while 
energy is redistributed at all wavelengths, including $\lambda > 912$ \AA.

In the studies mentioned above, ionising fluxes where calibrated as a function 
of effective temperature but not as a function of spectral type. Due to the cooler 
new effective temperature 
scales, such calibrations must be significantly different from previous relations based on 
unblanketed models. 
For astrophysical applications, the interesting quantity is in fact 
$Q_{i}$, the total number of ionising photons emitted per unit time related to $q_{i}$ by

\begin{equation}
Q_{i} = q_{i} \times 4 \pi R^{2},
\label{eq_Qi}
\end{equation}

\noindent with $R$ the stellar radius. With our calibrations \teff--ST and R--ST, we 
are now able to derive relations between spectral types and the total ionising 
fluxes $Q_{i}$. However, we have not considered the He{~\sc ii} ionising fluxes 
since they may be strongly affected by processes not included in our models 
(shocks, X-rays).

\begin{figure}
\centerline{\psfig{file=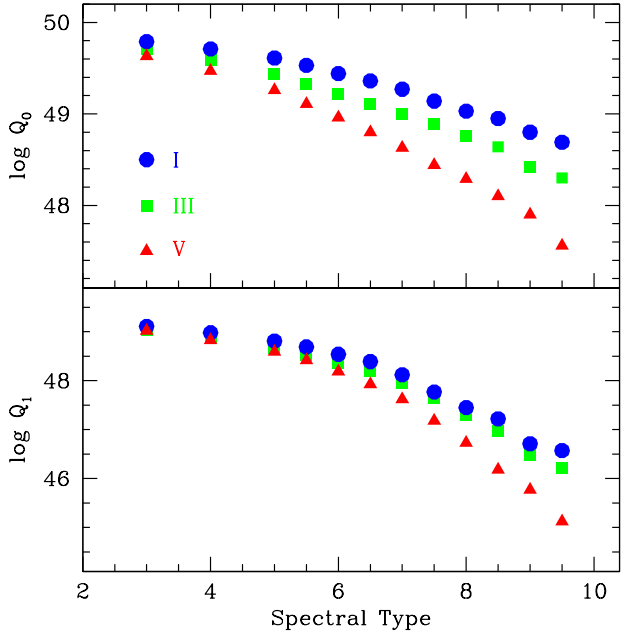,width=9cm}}
\caption{Ionising fluxes as a function of spectral type for dwarfs (triangles), 
giants (squares) and supergiants (circles) derived using the theoretical \teff\ scale.
}
\label{fig_Qi_ST}
\end{figure}

\begin{figure}
\centerline{\psfig{file=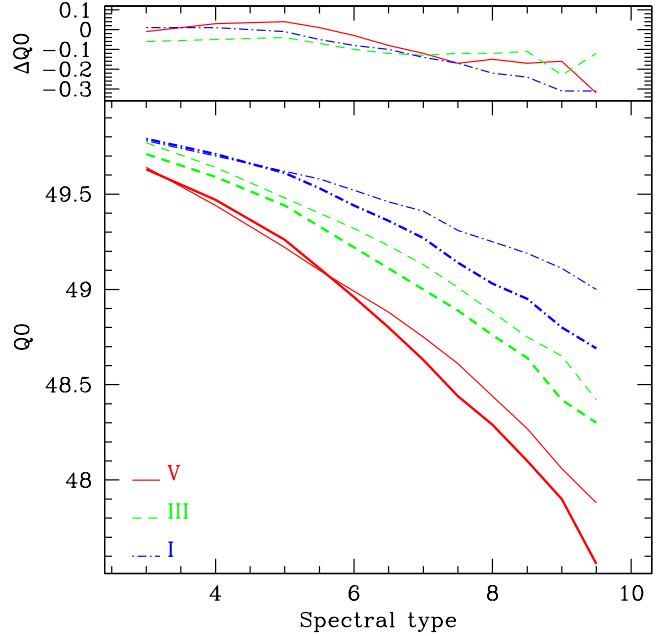,width=9cm}}
\caption{Lyman continuum photon fluxes derived using the theoretical (bold lines) and 
observational (thin lines) \teff\ scales. Solid lines are for dwarfs, dashed 
lines for giants and dot-dashed lines for supergiants. The upper panel shows 
the difference between theoretical and observational results.
}
\label{Q0_ST_diff}
\end{figure}

\begin{figure}
\centerline{\psfig{file=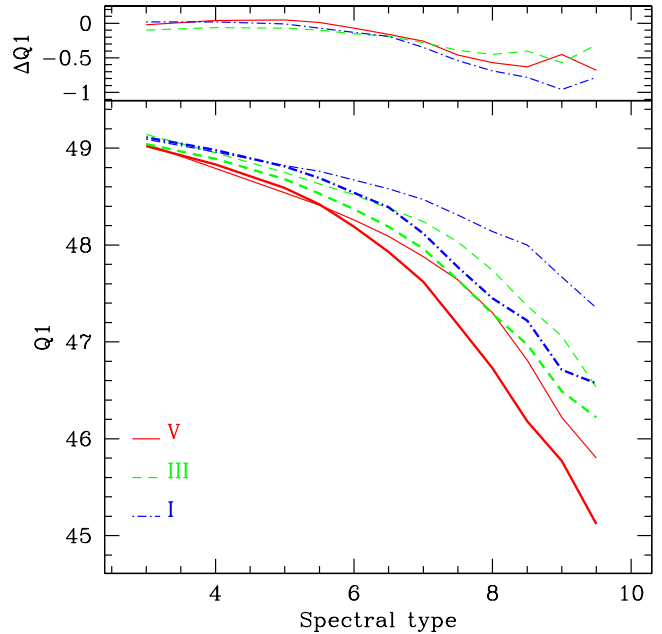,width=9cm}}
\caption{Same as Fig.\ \ref{Q0_ST_diff} but for He~{\sc i} ionising fluxes.
}
\label{Q1_ST_diff}
\end{figure}

First, we have to calibrate $q_{i}$. For that, we have proceeded as follows: we 
have built the function $q_{i}(\teff,\logg)$ using the same 2D interpolation 
method as for \teff\ (see Sect.\ \ref{teffscaletheo}) and then we have used the 
effective temperatures and gravities (relations \teff--ST and \logg--ST) 
for each luminosity class to derive the ionising fluxes of dwarfs, giants and 
supergiants. Once again, we have used both theoretical and observational \teff\
scales to investigate the effects of such relations on the final values of $q_{i}$
(see Tables \ref{tab_V} to \ref{tab_I_obs}).
Once obtained, these relations have been used with the relations R--ST 
to derive the calibration $Q_{i}$--spectral type. The results using the 
theoretical \teff\ scale are displayed in Fig.\ \ref{fig_Qi_ST}.
Figure \ref{Q0_ST_diff} 
and \ref{Q1_ST_diff} show the theoretical and observational $Q_{0}$'s and 
$Q_{1}$'s and the differences between them. Although for Lyman fluxes the 
differences remain relatively small (lower than 0.3 dex), using different 
\teff\ scales can have a deep impact on the He~{\sc i} ionising fluxes, 
especially for late type supergiants for which $Q_{1}$ may be changed by a 
factor 6!

\subsubsection{Comparisons with previous predictions}
\label{ion_flux_comparisons}

Due to the presence of metals in the model atmospheres, ionising fluxes of O stars 
are modified: new bound-free opacities increase the blocking of flux at 
short wavelengths and the numerous metallic lines change the very shape 
of the spectral energy distribution. 

Giveon et al.\ (\cite{giveon2}) and Morisset et al.\ (\cite{msbm04}) 
have shown that the new generation of atmosphere models including a non-LTE 
approach, winds and line-blanketing produced more accurate ionising fluxes 
compared to other previous less sophisticated models. They used the new 
SEDs to compute photoionisation models and to build excitation diagrams 
(defined by the ratio of mid-IR nebular lines of Ne, Ar and S). The 
direct comparison of such diagrams with observations of compact \hii\ 
regions with ISO showed the improvement of the new generation models.

\begin{figure}
\centerline{\psfig{file=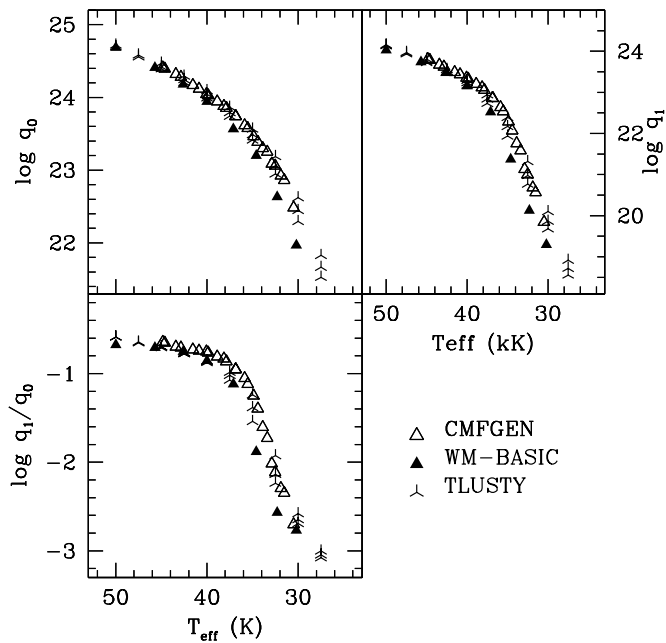,width=9cm}}
\caption{Comparison between CMFGEN, TLUSTY and 
WM-BASIC ionising fluxes for dwarfs. TLUSTY models are from Lanz \& Hubeny 
(\cite{ostar2002}) and WM-BASIC models from Smith et al.\ (\cite{smith}).
 Note that the latter have been estimated from the original SEDs and not 
from the SEDs re-binned for the inclusion in Starburst99. The CMFGEN fluxes 
are taken from Tables \ref{tab_V} to \ref{tab_I_obs}. 
}
\label{comp_qi_all_V}
\end{figure}

\begin{figure}
\centerline{\psfig{file=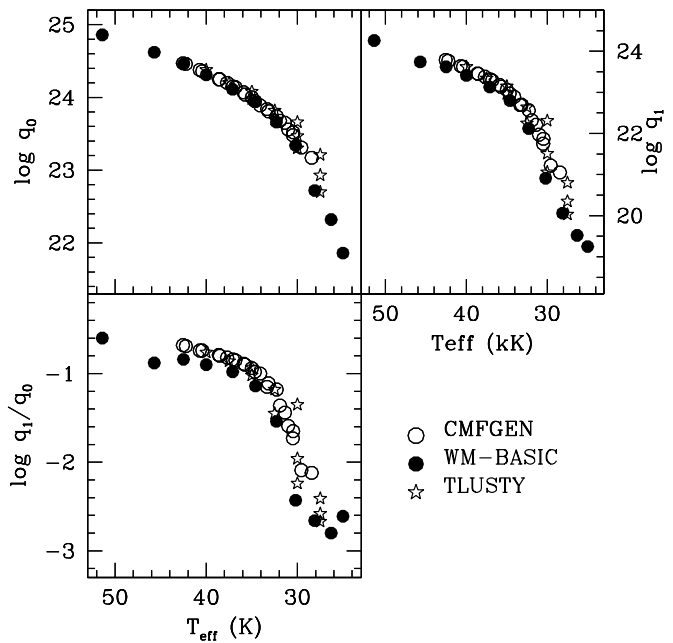,width=9cm}}
\caption{Same as Fig.\ \ref{comp_qi_all_V} for supergiants.
}
\label{comp_qi_all_I}
\end{figure}

Smith et al.\ (\cite{smith}) computed new ionisation fluxes for O stars
using  WM-BASIC (Pauldrach et al.\ 
\cite{wmbasic01}). Lanz \& Hubeny 
(\cite{ostar2002}) built a grid of plane parallel line-blanketed non-LTE 
models with TLUSTY and provided H and HeI ionising fluxes. 
Figures \ref{comp_qi_all_V} and \ref{comp_qi_all_I} show the comparison between 
our ionising fluxes 
as a function of \teff\ for dwarfs and supergiants with those of 
Smith et al.\ (\cite{smith}) 
and the TLUSTY grid. Concerning H ionising fluxes, one notes that 
CMFGEN and TLUSTY agree very well (within 0.1 dex). WM-BASIC models also 
show similar fluxes for supergiants and for the hottest dwarfs, while \qo\ are 
slightly lower (up to 0.3 dex) for dwarfs with \teff\ $<$ 40000 K. The 
different gravity between our CMFGEN dwarfs ($\logg \sim 3.9$) and the 
WM-BASIC dwarfs ($\logg = 4.0$) may be partly responsible for this 
difference. However, since the difference in \qo\ are equivalent to a 
shift in \teff\ by less than 2000 K (see Fig.\ \ref{comp_qi_all_V}), 
we may speculate that a different treatment of line-blanketing between 
CMFGEN and WM-BASIC may be the main reason for the discrepancy. 
Concerning the HeI ionising fluxes, CMFGEN and TLUSTY are also in good 
agreement, which may be surprising given that TLUSTY models do not include 
winds. However, the largest wind effects are seen in the HeII continuum 
(Hillier \cite{hil87a}, \cite{hil87b}, Gabler et al.\ \cite{gabler89}, 
Schaerer \& de Koter \cite{costar3}). 
Again, the WM-BASIC HeI ionising fluxes are lower for cool dwarfs ($\sim$ 
0.3 dex) and supergiants ($\sim$ 0.1 dex). This results in softer spectra 
for WM-BASIC models compared to both CMFGEN and TLUSTY models for late 
dwarfs and supergiants. 
The exact reasons for this difference
leading to an apparent relative overestimate of line blanketing remain 
unclear but may be likely 
attributed to different methods used to treat line-blanketing (opacity 
sampling versus direct inclusion with super-level). 
The behaviour of several 
atmosphere models including CMFGEN, TLUSTY and WM-BASIC have been 
discussed extensively by Morisset et al.\ (\cite{msbm04}). 
In short, the CMFGEN and WM-BASIC models do a reasonable job when
compared to observed fine structure line ratios of Galactic \hii\ regions.
However, at present the nebular analysis does not allow to favour
CMFGEN or WM-BASIC models. New tailored studies of individual nebulae
(following e.g.\ the steps of Oey et al.\ \cite{oey00})
using these fully blanketed non-LTE models would be required to provide
more strongest tests.

\begin{figure}
\centerline{\psfig{file=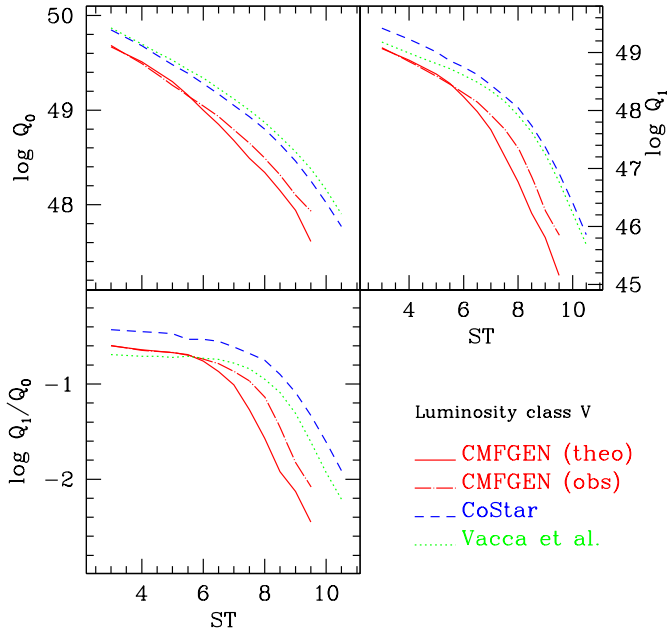,width=9cm}}
\caption{Comparison between the present ionising fluxes of dwarfs 
and the results 
of Vacca et al.\ \cite{vacca} and Schaerer \& de Koter (\cite{costar3}, 
CoStar models).}
\label{comp_Qi_V}
\end{figure}

\begin{figure}
\centerline{\psfig{file=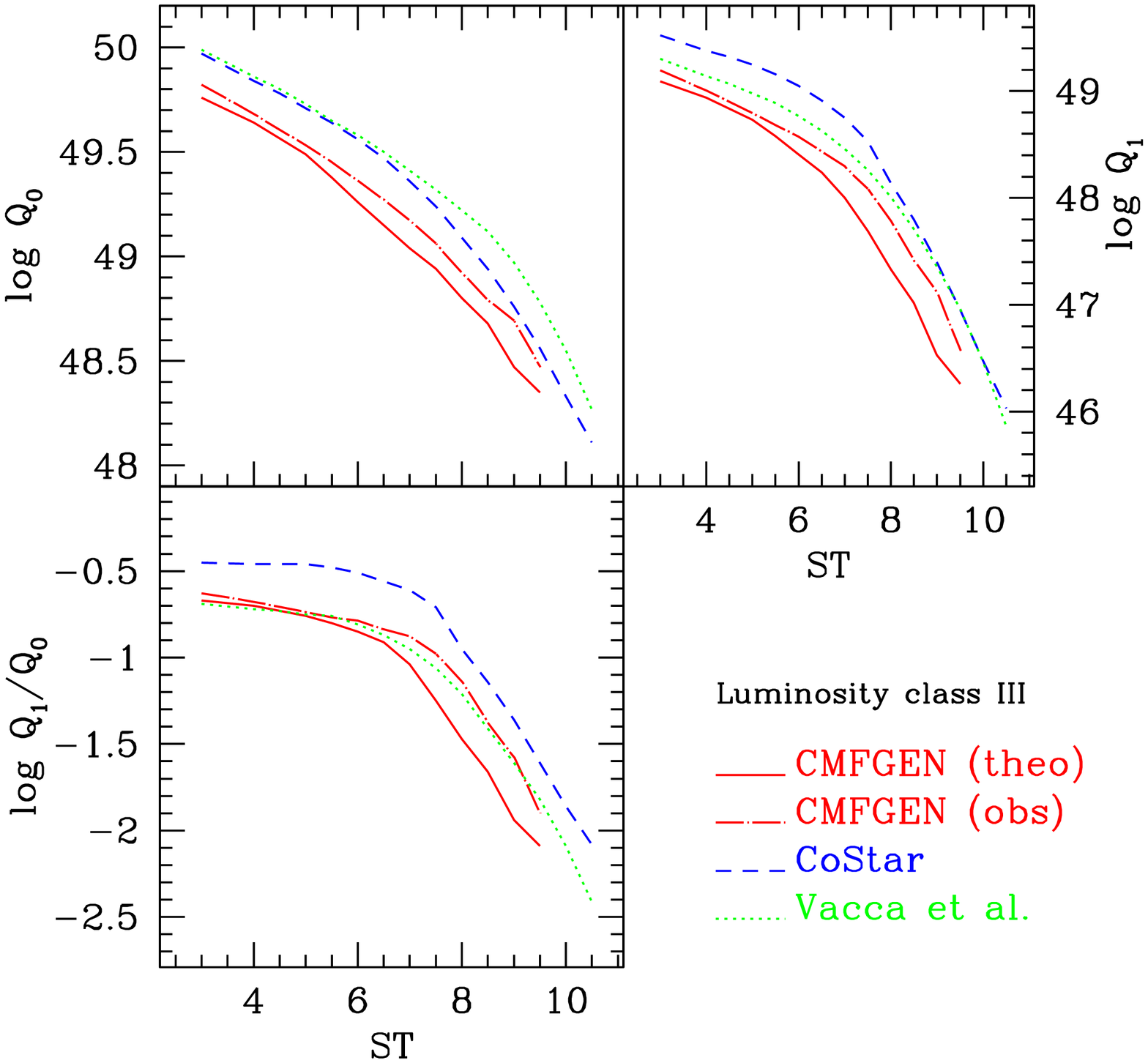,width=9cm}}
\caption{Same as Fig.\ \ref{comp_Qi_V} for giants.}
\label{comp_Qi_III}
\end{figure}

\begin{figure}
\centerline{\psfig{file=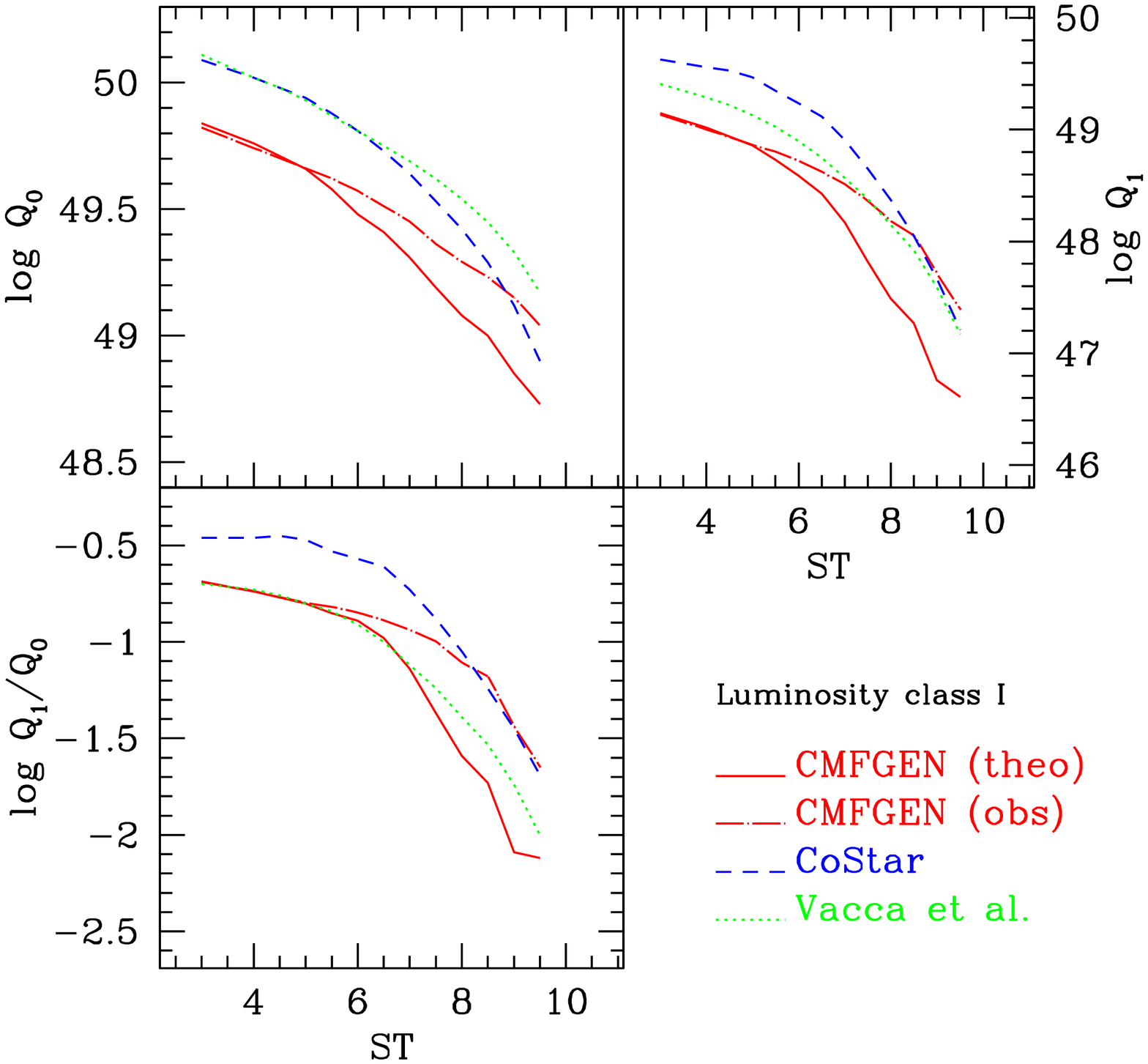,width=9cm}}
\caption{Same as Fig.\ \ref{comp_Qi_V} for supergiants.}
\label{comp_Qi_I}
\end{figure}

The relations $q_{i}(T_{\rm eff})$ allow a direct comparison of atmosphere 
models, but the quantities of interest for a number of studies (\hii\ regions, 
nebular analysis...) are usually the integrated ionising fluxes $Q_{i}$ 
calibrated as function of spectral type. Such relations ($Q_{i}$--$ST$) 
depend not only on the model ingredients, but also on the effective 
temperature scale and on the R--ST relation.    
Figures \ref{comp_Qi_V}, 
\ref{comp_Qi_III} and \ref{comp_Qi_I} show how our ionising fluxes 
-- spectral type relations derived using both the theoretical and 
observational effective temperature scales
compare to the 
previous calibrations of Vacca et al.\ (\cite{vacca}) and Schaerer \& de 
Koter (\cite{costar3}, CoStar models). 

Depending on which \teff\ scale is used, two kinds of conclusions 
can be drawn.
First, the theoretical ionising fluxes are lower than in 
the previous calibrations. 
Compared to the results of Vacca et al.\ (\cite{vacca}), our $Q_{0}$'s 
are 0.20 to 0.80 dex lower for dwarfs, 0.25 to 0.55 lower for giants and 
0.30 to 0.55 dex lower for supergiants. 
For $Q_{1}$, the reduction is 0.1 to 1.6(!) dex for LC V, 0.2 to 0.8 
dex for LC III and 0.25 to 0.8 dex for LC I.
Such reductions are due to two distinct effects:
first, the better treatment of line-blanketing compared to the CoStar 
models and the inclusion of both metals and winds compared to the 
results of Vacca et al.\ (\cite{vacca}), which modifies the relations 
$q_{i}(T_{\rm eff})$; second, new relations \teff--ST and R--ST 
are used to compute $Q_{i}$. These effects render 
the CMFGEN spectra softer than the CoStar spectra 
($\log Q_{1}/Q_{0}$ lower by $\sim$ 0.2 to 1.0 dex). The Vacca et al.\ 
(\cite{vacca}) predictions are usually harder than the CMFGEN SEDs, 
except for the earliest spectral types. 

Now if we use the \teff\ scales derived from the results of detailed 
spectroscopic analysis of individual massive stars, the ionising fluxes 
of stars with spectral types earlier than O6 are basically unchanged 
compared to the results based on the ``theoretical'' \teff\ scale. 
For later spectral types the reduction of $Q_{i}$ for a given 
ST is much weaker, $Q_{0}$ and $Q_{1}$ being essentially similar to the Costar 
and Vacca et al.\ (\cite{vacca}) results for late supergiants. 
Quantitatively, the reduction of $Q_{0}$ is 0.25 to 0.50 dex (0.20 to 
0.35 dex) for dwarfs (giants and supergiants). 
$Q_{1}$ is much less reduced with the observational scale 
(0.15 to 1.0 dex for dwarfs and 0.15 to 0.45 dex for giants) and are in 
fact similar $-$within 0.23 dex$-$ to the Vacca et al.\ (\cite{vacca}) values for late 
supergiants.   

What do we learn from this? The obvious conclusion is that the choice 
of the underlying effective temperature scale is important to determine the ionising 
fluxes for a given spectral type and luminosity class.
As mentioned in Sect.\ \ref{teffscales}, both the theoretical 
and observational scales have their advantages and drawbacks. At present, we 
cannot say that one is more relevant than the other, which leaves us 
with an uncertainty on the ionising fluxes. However, both types of \teff\ 
- scales are significantly cooler than the ones based on plane-parallel 
H He models. 
Besides this, it is also worth noting that ionising fluxes are very 
sensitive to \teff\ (especially at low values) so that the difference 
we find between calibrations of $Q_{i}$s using different \teff\ - scales 
also puts forward the fact that the natural spread in \teff\ for a 
given spectral type (see Figs. 
\ref{Teff_ST_V} to \ref{Teff_ST_I}) leads to an uncertainty on $Q_{i}$s. 
And the error on the spectral 
type (of the order 0.5 unit) also translates to an error on the 
ionising fluxes (which amounts to a factor of  $\sim$ 2 on $Q_{1}$ for 
late spectral types). 
Thus, the present results are an improvement over previous 
analysis and calibrations, but it should be reminded that they still 
suffer from uncertainties.

%----------------------------------------------------------------------------
%                                 Tables
%----------------------------------------------------------------------------

\begin{table*}
\caption{Stellar parameters as a function of spectral types for luminosity 
class V stars obtained with the theoretical \teff\ scales.
}
\label{tab_V}
\center
\begin{tabular}{l|llllllllllll}
\hline 
ST & \teff\ & $\log g_{spec}$ & \mv\ & BC & \lL & R & M$_{spec}$ & \qo\ & \qi\ & \Qo\ & \Qi\ & \\
 & [K] & [cm s$^{-2}$] &  &  &  & [\rsun] & [\msun] & [cm$^{-2}$ s$^{-1}$] & [cm$^{-2}$ s$^{-1}$] & [s$^{-1}$] & [s$^{-1}$] &\\ 
\hline
3    & 44616 & 3.92 & -5.79 & -4.03 & 5.83 & 13.84 & 58.34 & 24.56 & 23.96 & 49.63 & 49.02 & \\
4    & 43419 & 3.92 & -5.50 & -3.95 & 5.68 & 12.31 & 46.16 & 24.50 & 23.86 & 49.47 & 48.83 & \\
5    & 41540 & 3.92 & -5.21 & -3.82 & 5.51 & 11.08 & 37.28 & 24.38 & 23.71 & 49.26 & 48.59 & \\
5.5  & 40062 & 3.92 & -5.07 & -3.71 & 5.41 & 10.61 & 34.17 & 24.28 & 23.59 & 49.11 & 48.42 & \\
6    & 38151 & 3.92 & -4.92 & -3.57 & 5.30 & 10.23 & 31.73 & 24.15 & 23.39 & 48.96 & 48.19 & \\
6.5  & 36826 & 3.92 & -4.77 & -3.47 & 5.20 & 9.79  & 29.02 & 24.04 & 23.17 & 48.80 & 47.93 & \\
7    & 35531 & 3.92 & -4.63 & -3.36 & 5.10 & 9.37  & 26.52 & 23.91 & 22.89 & 48.63 & 47.62 & \\
7.5  & 34419 & 3.92 & -4.48 & -3.27 & 5.00 & 8.94  & 24.15 & 23.76 & 22.49 & 48.44 & 47.18 & \\
8    & 33383 & 3.92 & -4.34 & -3.18 & 4.90 & 8.52  & 21.95 & 23.65 & 22.08 & 48.29 & 46.73 & \\
8.5  & 32522 & 3.92 & -4.19 & -3.10 & 4.82 & 8.11  & 19.82 & 23.50 & 21.58 & 48.10 & 46.18 & \\
9    & 31524 & 3.92 & -4.05 & -3.01 & 4.72 & 7.73  & 18.03 & 23.34 & 21.21 & 47.90 & 45.77 & \\
9.5  & 30488 & 3.92 & -3.90 & -2.91 & 4.62 & 7.39  & 16.46 & 23.04 & 20.59 & 47.56 & 45.12 & \\
\hline 
\end{tabular}
\end{table*}

\begin{table*}
\caption{Same as Table \ref{tab_V} for luminosity class III stars.}
\label{tab_III}
\center
\begin{tabular}{l|llllllllllll}
\hline 
ST & \teff\ & $\log g_{spec}$ & \mv\ & BC & \lL & R & M$_{spec}$ & \qo\ & \qi\ & \Qo\ & \Qi\ & \\
 & [K] & [cm s$^{-2}$] &  &  &  & [\rsun] & [\msun] & [cm$^{-2}$ s$^{-1}$] & [cm$^{-2}$ s$^{-1}$] & [s$^{-1}$] & [s$^{-1}$] &\\ 
\hline
3    & 42942 & 3.77 & -6.13 & -3.92 & 5.92 & 16.57 & 58.62 & 24.49 & 23.82 & 49.71 & 49.04 & \\
4    & 41486 & 3.73 & -5.98 & -3.82 & 5.82 & 15.83 & 48.80 & 24.41 & 23.71 & 49.59 & 48.89 & \\
5    & 39507 & 3.69 & -5.84 & -3.67 & 5.70 & 15.26 & 41.48 & 24.29 & 23.53 & 49.44 & 48.68 & \\
5.5  & 38003 & 3.67 & -5.76 & -3.56 & 5.63 & 15.13 & 38.92 & 24.19 & 23.39 & 49.33 & 48.53 & \\
6    & 36673 & 3.65 & -5.69 & -3.45 & 5.56 & 14.97 & 36.38 & 24.08 & 23.23 & 49.22 & 48.37 & \\
6.5  & 35644 & 3.63 & -5.62 & -3.37 & 5.49 & 14.74 & 33.68 & 23.99 & 23.07 & 49.11 & 48.19 & \\
7    & 34638 & 3.61 & -5.54 & -3.28 & 5.43 & 14.51 & 31.17 & 23.89 & 22.85 & 49.00 & 47.96 & \\
7.5  & 33487 & 3.59 & -5.47 & -3.18 & 5.36 & 14.34 & 29.06 & 23.79 & 22.55 & 48.89 & 47.64 & \\
8    & 32573 & 3.57 & -5.40 & -3.10 & 5.30 & 14.11 & 26.89 & 23.68 & 22.21 & 48.76 & 47.29 & \\
8.5  & 31689 & 3.55 & -5.32 & -3.02 & 5.24 & 13.88 & 24.84 & 23.57 & 21.90 & 48.64 & 46.97 & \\
9    & 30737 & 3.53 & -5.25 & -2.93 & 5.17 & 13.69 & 23.07 & 23.37 & 21.43 & 48.42 & 46.49 & \\
9.5  & 30231 & 3.51 & -5.18 & -2.88 & 5.12 & 13.37 & 21.04 & 23.27 & 21.18 & 48.30 & 46.22 & \\ 
\hline 
\end{tabular}
\end{table*}

\begin{table*}
\caption{Same as Table \ref{tab_V} for luminosity class I stars.}
\label{tab_I}
\center
\begin{tabular}{l|llllllllllll}
\hline 
ST & \teff\ & $\log g_{spec}$ & \mv\ & BC & \lL & R & M$_{spec}$ & \qo\ & \qi\ & \Qo\ & \Qi\ & \\
 & [K] & [cm s$^{-2}$] &  &  &  & [\rsun] & [\msun] & [cm$^{-2}$ s$^{-1}$] & [cm$^{-2}$ s$^{-1}$] & [s$^{-1}$] & [s$^{-1}$] &\\ 
\hline
3    & 42551 & 3.73 & -6.35 & -3.89 & 6.00 & 18.47 & 66.89 & 24.47 & 23.79 & 49.79 & 49.11 & \\
4    & 40702 & 3.65 & -6.34 & -3.76 & 5.94 & 18.91 & 58.03 & 24.38 & 23.64 & 49.71 & 48.98 & \\
5    & 38520 & 3.57 & -6.33 & -3.60 & 5.87 & 19.48 & 50.87 & 24.25 & 23.45 & 49.61 & 48.81 & \\
5.5  & 37070 & 3.52 & -6.33 & -3.48 & 5.82 & 19.92 & 48.29 & 24.15 & 23.31 & 49.53 & 48.69 & \\
6    & 35747 & 3.48 & -6.32 & -3.38 & 5.78 & 20.33 & 45.78 & 24.04 & 23.14 & 49.44 & 48.54 & \\
6.5  & 34654 & 3.44 & -6.31 & -3.29 & 5.74 & 20.68 & 43.10 & 23.95 & 22.97 & 49.36 & 48.39 & \\
7    & 33326 & 3.40 & -6.31 & -3.17 & 5.69 & 21.14 & 40.91 & 23.84 & 22.69 & 49.27 & 48.12 & \\
7.5  & 31913 & 3.36 & -6.30 & -3.04 & 5.64 & 21.69 & 39.17 & 23.68 & 22.32 & 49.14 & 47.77 & \\
8    & 31009 & 3.32 & -6.30 & -2.96 & 5.60 & 22.03 & 36.77 & 23.56 & 21.97 & 49.03 & 47.45 & \\
8.5  & 30504 & 3.28 & -6.29 & -2.91 & 5.58 & 22.20 & 33.90 & 23.48 & 21.75 & 48.95 & 47.22 & \\
9    & 29569 & 3.23 & -6.29 & -2.82 & 5.54 & 22.60 & 31.95 & 23.31 & 21.22 & 48.80 & 46.71 & \\
9.5  & 28430 & 3.19 & -6.28 & -2.70 & 5.49 & 23.11 & 30.41 & 23.17 & 21.05 & 48.69 & 46.57 & \\
\hline 
\end{tabular}
\end{table*}

\begin{table*}
\caption{Stellar parameters as a function of spectral types for luminosity 
class V stars obtained with the observational \teff\ scales.
}
\label{tab_V_obs}
\center
\begin{tabular}{l|llllllllllll}
\hline 
ST & \teff\ & $\log g_{spec}$ & \mv\ & BC & \lL & R & M$_{spec}$ & \qo\ & \qi\ & \Qo\ & \Qi\ & \\
 & [K] & [cm s$^{-2}$] &  &  &  & [\rsun] & [\msun] & [cm$^{-2}$ s$^{-1}$] & [cm$^{-2}$ s$^{-1}$] & [s$^{-1}$] & [s$^{-1}$] &\\ 
\hline
3    & 44852 & 3.92 & -5.79 & -4.05 & 5.84 & 13.80 & 57.95 & 24.57 & 23.98 & 49.64 & 49.04 & \\
4    & 42857 & 3.92 & -5.50 & -3.91 & 5.67 & 12.42 & 46.94 & 24.47 & 23.82 & 49.44 & 48.79 & \\
5    & 40862 & 3.92 & -5.21 & -3.77 & 5.49 & 11.20 & 38.08 & 24.34 & 23.66 & 49.22 & 48.54 & \\
5.5  & 39865 & 3.92 & -5.07 & -3.70 & 5.41 & 10.64 & 34.39 & 24.27 & 23.57 & 49.10 & 48.41 & \\
6    & 38867 & 3.92 & -4.92 & -3.62 & 5.32 & 10.11 & 30.98 & 24.20 & 23.46 & 48.99 & 48.26 & \\
6.5  & 37870 & 3.92 & -4.77 & -3.55 & 5.23 & 9.61  & 28.00 & 24.13 & 23.34 & 48.88 & 48.09 & \\
7    & 36872 & 3.92 & -4.63 & -3.47 & 5.14 & 9.15  & 25.29 & 24.04 & 23.17 & 48.75 & 47.88 & \\
7.5  & 35874 & 3.92 & -4.48 & -3.39 & 5.05 & 8.70  & 22.90 & 23.94 & 22.98 & 48.61 & 47.64 & \\
8    & 34877 & 3.92 & -4.34 & -3.30 & 4.96 & 8.29  & 20.76 & 23.82 & 22.68 & 48.44 & 47.30 & \\
8.5  & 33879 & 3.92 & -4.19 & -3.22 & 4.86 & 7.90  & 18.80 & 23.69 & 22.23 & 48.27 & 46.81 & \\
9    & 32882 & 3.92 & -4.05 & -3.13 & 4.77 & 7.53  & 17.08 & 23.52 & 21.69 & 48.06 & 46.22 & \\
9.5  & 31884 & 3.92 & -3.90 & -3.04 & 4.68 & 7.18  & 15.55 & 23.39 & 21.31 & 47.88 & 45.80 & \\
\hline 
\end{tabular}
\end{table*}

\begin{table*}
\caption{Same as Table \ref{tab_V_obs} for luminosity class III stars.}
\label{tab_III_obs}
\center
\begin{tabular}{l|llllllllllll}
\hline 
ST & \teff\ & $\log g_{spec}$ & \mv\ & BC & \lL & R & M$_{spec}$ & \qo\ & \qi\ & \Qo\ & \Qi\ & \\
 & [K] & [cm s$^{-2}$] &  &  &  & [\rsun] & [\msun] & [cm$^{-2}$ s$^{-1}$] & [cm$^{-2}$ s$^{-1}$] & [s$^{-1}$] & [s$^{-1}$] &\\ 
\hline
3    & 44537 & 3.77 & -6.13 & -4.03 & 5.96 & 16.19 & 55.95 & 24.57 & 23.94 & 49.77 & 49.14 & \\
4    & 42422 & 3.73 & -5.98 & -3.88 & 5.85 & 15.61 & 47.43 & 24.47 & 23.78 & 49.64 & 48.95 & \\
5    & 40307 & 3.69 & -5.84 & -3.73 & 5.73 & 15.07 & 40.43 & 24.34 & 23.61 & 49.48 & 48.75 & \\
5.5  & 39249 & 3.67 & -5.76 & -3.65 & 5.67 & 14.82 & 37.35 & 24.28 & 23.51 & 49.40 & 48.63 & \\
6    & 38192 & 3.65 & -5.69 & -3.57 & 5.61 & 14.58 & 34.53 & 24.21 & 23.41 & 49.32 & 48.52 & \\
6.5  & 37134 & 3.63 & -5.62 & -3.49 & 5.54 & 14.36 & 31.96 & 24.13 & 23.28 & 49.23 & 48.38 & \\
7    & 36077 & 3.61 & -5.54 & -3.40 & 5.48 & 14.14 & 29.59 & 24.04 & 23.16 & 49.13 & 48.24 & \\
7.5  & 35019 & 3.59 & -5.47 & -3.32 & 5.42 & 13.93 & 27.45 & 23.94 & 22.96 & 49.01 & 48.03 & \\
8    & 33961 & 3.57 & -5.40 & -3.23 & 5.35 & 13.74 & 25.49 & 23.82 & 22.68 & 48.88 & 47.74 & \\
8.5  & 32904 & 3.55 & -5.32 & -3.13 & 5.28 & 13.55 & 23.68 & 23.70 & 22.32 & 48.75 & 47.37 & \\
9    & 31846 & 3.53 & -5.25 & -3.04 & 5.21 & 13.38 & 22.04 & 23.61 & 22.02 & 48.65 & 47.06 & \\
9.5  & 30789 & 3.51 & -5.18 & -2.94 & 5.15 & 13.22 & 20.55 & 23.39 & 21.50 & 48.42 & 46.53 & \\
\hline 
\end{tabular}
\end{table*}

\begin{table*}
\caption{Same as Table \ref{tab_V_obs} for luminosity class I stars.}
\label{tab_I_obs}
\center
\begin{tabular}{l|llllllllllll}
\hline 
ST & \teff\ & $\log g_{spec}$ & \mv\ & BC & \lL & R & M$_{spec}$ & \qo\ & \qi\ & \Qo\ & \Qi\ & \\
 & [K] & [cm s$^{-2}$] &  &  &  & [\rsun] & [\msun] & [cm$^{-2}$ s$^{-1}$] & [cm$^{-2}$ s$^{-1}$] & [s$^{-1}$] & [s$^{-1}$] &\\ 
\hline
3    & 42233 & 3.73 & -6.35 & -3.87 & 5.99 & 18.56 & 67.53 & 24.46 & 23.77 & 49.78 & 49.09 & \\
4    & 40422 & 3.65 & -6.34 & -3.74 & 5.93 & 18.99 & 58.54 & 24.36 & 23.62 & 49.70 & 48.96 & \\
5    & 38612 & 3.57 & -6.33 & -3.61 & 5.87 & 19.45 & 50.72 & 24.25 & 23.46 & 49.62 & 48.82 & \\
5.5  & 37706 & 3.52 & -6.33 & -3.54 & 5.84 & 19.70 & 47.25 & 24.20 & 23.38 & 49.58 & 48.76 & \\
6    & 36801 & 3.48 & -6.32 & -3.46 & 5.81 & 19.95 & 44.10 & 24.14 & 23.29 & 49.52 & 48.67 & \\
6.5  & 35895 & 3.44 & -6.31 & -3.39 & 5.78 & 20.22 & 41.20 & 24.07 & 23.18 & 49.46 & 48.58 & \\
7    & 34990 & 3.40 & -6.31 & -3.31 & 5.75 & 20.49 & 38.44 & 24.00 & 23.06 & 49.41 & 48.47 & \\
7.5  & 34084 & 3.36 & -6.30 & -3.24 & 5.72 & 20.79 & 36.00 & 23.89 & 22.89 & 49.31 & 48.31 & \\
8    & 33179 & 3.32 & -6.30 & -3.16 & 5.68 & 21.10 & 33.72 & 23.81 & 22.70 & 49.25 & 48.14 & \\
8.5  & 32274 & 3.28 & -6.29 & -3.08 & 5.65 & 21.41 & 31.54 & 23.74 & 22.56 & 49.19 & 48.00 & \\
9    & 31368 & 3.23 & -6.29 & -2.99 & 5.61 & 21.76 & 29.63 & 23.65 & 22.21 & 49.11 & 47.67 & \\
9.5  & 30463 & 3.19x & -6.28 & -2.91 & 5.57 & 22.11 & 27.83 & 23.52 & 21.87 & 49.00 & 47.35 & \\
\hline 
\end{tabular}
\end{table*}

%%%%%%%%%%%%%%%%%%%%%%%%%%%%%%%%%%%%%%%%%%%%%%%%%%%%%%%%%%%%%%%%%%%%%%%%
%----------------------------------------------------------------------
%
%         Discussion
%
%----------------------------------------------------------------------
\section{Discussion: metallicity effects}
\label{disc_Z}

The present study is restricted to the case of a solar metallicity 
for atmosphere models.
First, we want to highlight that we have used a solar composition for all 
luminosity classes. This assumption is certainly good for dwarfs, but may 
be too crude for supergiants which usually 
show hints of He enrichment and modified CNO abundances (e.g. Herrero et 
al.\ \cite{her92}, Walborn et al.\ \cite{walborn04}). These changes are 
likely different from star to star since they depend on rotational velocities 
(Meynet \& Maeder \cite{mm00}). Hence, non solar He abundance may slightly 
affect the strength of the classification lines and introduce a dispersion in ST
for a given \teff. Also, although Iron dominates the blanketing effect in terms 
of lines, CNO elements are important for the blocking of radiation through their 
bound-free transitions usually close to the He~{\sc II} edge so that, although the 
sum of C, N and O abundances are constant during CNO processing, a change in the 
abundance of each element may modify the global blanketing effect. Such an effect 
should be further studied by new analyses of supergiants.
%However, non solar CNO
%abundances should not modify our results since most of the blanketing effect 
%is due to Iron. Moreover the sum of C, N and O abundances remains constant 
%during CNO processing.    
Second, the present work does not address the important question of the 
dependence of the fundamental parameters of O stars on global metallicity. 
Nonetheless, several indications of the general trend exist. 

From the modelling side, Kudritzki \cite{kud02} and Mokiem et 
al.\ (\cite{mokiem}) 
have investigated the effect of a change of the metal content on the 
spectral energy distribution of O dwarfs using CMFGEN models. They 
found that H ionising fluxes are essentially unchanged when Z is varied
 between twice and one tenth the solar content. They argue that the 
redistribution of the flux blocked by metals at short wavelengths takes 
place within the Lyman continuum, which explains the observed behaviour. 
However, they show that the SEDs are strongly modified below $\sim$ 450 
\AA, spectra being softer at higher metallicity (see also Sect.\ 
\ref{ion_flux_calib}). Morisset et al.\ 
(\cite{msbm04}) have computed various WM-BASIC models at different 
metallicities and showed how Z affected the strength of mid-IR 
nebular lines emitted in compact \hii\ regions. The softening of the SEDs 
when metallicity increases is crucial to understand the behaviour of 
observed excitations sequences.
  
Concerning the effect of metallicity on \teff\ scales, Mokiem et al.\ 
(\cite{mokiem}) have shown that for a given \teff, spectral types 
vary within one subclass when Z is decreased from 2 to 0.1 $Z_{\odot}$. 
This boils down to a higher effective temperature at low metallicity than 
at solar metallicity.
We reach the same conclusion from 
the study of several test models for dwarfs with Z = 1/8 $Z_{\odot}$ 
(see Martins \cite{thesis}). We estimate that the reduction of the \teff\ 
scale (compared to pure H He results) at this metallicity is roughly 
half the reduction obtained at solar metallicity (see also Martins et 
al.\ \cite{teffscale}).

Observationally, Massey et al. (\cite{massey04}) derived effective 
temperatures of a sample of O stars in the Magellanic Clouds by means 
of models computed with the code 
FASTWIND (Santolaya-Rey et al. \cite{fast97}).
They found lower \teff\ than Vacca et al.\ (\cite{vacca}) 
but higher than those of Galactic O stars, in good agreement with
our results. They estimate that effective temperatures 
of early to mid O type MC objects are 3000 to 4000 K hotter than 
Galactic counterparts. Previously, Crowther et al.\ (\cite{paul02}) 
computed CMFGEN models and
showed that extreme O supergiants in the Magellanic Clouds were cooler 
by 5000 to 7500 K compared to the Vacca et al.\ (\cite{vacca}) calibration. 
Bouret et al.\ (\cite{jc03}) also confirmed the reduction of effective 
temperatures in SMC O dwarfs using CMFGEN models. However, they 
derived temperatures in agreement our \teff\ scale for Galactic stars, 
which may be surprising given the lower metallicity of the SMC. 
Even more recently, Heap et al.\ (\cite{heap05}) submitted a paper 
in which they derive the stellar properties of a large sample of SMC stars. 
Again, the effective temperatures are not inconsistent with those of Galactic 
stars. They also provide bolometric corrections and ionising fluxes which are 
in better agreement with our new calibration than with the older pure H He 
plane-parallel ones. Given the dispersion in the data points, a conclusion as 
regards the metallicity dependence is not possible.

The above discussion shows that although hints on the metallicity 
dependence of stellar parameters of O stars exists, a quantitative 
and systematic study remains to be carried out.

%%%%%%%%%%%%%%%%%%%%%%%%%%%%%%%%%%%%%%%%%%%%%%%%%%%%%%%%%%%%%%%%%%%%%%%%
%----------------------------------------------------------------------
%
%         Conclusion
%
%----------------------------------------------------------------------
\section{Concluding remarks}
\label{conc}

We have presented new calibrations of stellar parameters of O stars as 
a function of spectral type based on atmosphere models computed with
the code CMFGEN (Hillier \& Miller \cite{hm98}). We have build a grid
of models spanning the range of spectral types and luminosity classes
of O stars from which calibrations have been derived through 2D
interpolations. The main improvement of such relations over previous
ones is the inclusion of line-blanketing and winds in the non-LTE atmosphere
models.
Our main results are the following:

\begin{itemize}

\item[$\diamond$] We have derived two types of effective temperature scales: 
a theoretical 
one based uniquely on our grid of models (approach similar to paper I) 
and an observational one from the results of detailed spectroscopic 
analysis of individual O stars with line blanketed non-LTE models 
including winds. 
We confirm the now well established fact that line-blanketing leads to 
cooler effective temperature scales compared to the widely used
relations of Vacca et al. (\cite{vacca}) based on plane-parallel pure
H He models. The adoption of a better photospheric structure in our
models leads to slightly cooler ($\sim$ 500 K) theoretical effective temperatures
compared to our first results (Martins et al.\ \cite{teffscale}).
Our theoretical \teff--ST relations are cooler by 2000 to 8000 K --- being larger 
for earlier spectral types and lower luminosity classes --- compared 
to Vacca et al.\ (\cite{vacca}) calibrations. 
The theoretical \teff\ scales are consistent with observational 
relations for early type dwarfs and supergiants and are slightly cooler 
for late type spectral types (by up to 2000 K for supergiants). This may 
be due to too
high mass loss rates in our models.
For giants, the theoretical relation is systematically cooler by $\sim$ 1000 K.
We estimate the uncertainty of our theoretical 
effective temperatures to be between 1000 and 2000 K for a given
spectral type. This is due to the intrinsic dispersion of temperatures for O
stars of a given spectral type and to the uncertainties of
the models (hydrodynamic structure, adopted parameters). 

\item[$\diamond$] Bolometric corrections are reduced compared to the 
calibrations of Vacca et al.\ (\cite{vacca}), the reduction being larger 
for supergiants and for the earliest spectral types. 
BC's estimated using the theoretical \teff\ scales are 
0.40 to 0.50 (resp.\ 0.35 to 0.60, 0.40 to 0.65) mag lower for dwarfs (resp. 
giants, supergiants), while those estimated using the observational 
relations are 0.30 to 0.50 (resp.\ 0.30 to 0.50, 0.20 to 0.65) mag lower for 
luminosity class V (resp.\ III, I). 

\item[$\diamond$] Luminosities are reduced by 0.20 to 0.35 dex for dwarfs, 
0.25 dex for giants and 0.25 to 0.35 dex for supergiants compared to the calibrations
of Vacca et al.\ (\cite{vacca}) with little difference ($\lesssim$ 0.1 dex) 
between results using theoretical and observational relations.  The 
reduction of luminosity is independent 
of spectral type for giants and supergiants and is slightly larger for late type 
than for early type dwarfs.

\item[$\diamond$] For a given spectral type, ionising fluxes are 
reduced due to the effect of line-blanketing
on the SED (blocking of flux). Compared to the Vacca et al.\
(\cite{vacca}) values and using our theoretical \teff\ scale, we find 
integrated Lyman ionising fluxes ($Q_{0}$) 0.20 to 0.80 dex lower for dwarfs,
the larger difference being at late spectral types.
The reduction is 0.25 to 0.55 dex (resp. 
0.30 to 0.55 dex) for giants (resp. supergiants). If we use the observational 
effective temperature scale, the reduction are 0.25 to 0.50 dex for dwarfs 
and 0.20 to 0.35 dex for giants and supergiants. He~{\sc i} ionising fluxes 
are also reduced, the difference between theoretical and observational 
results being larger.

For a given
\teff, our $q_{i}$'s agree well with the TLUSTY grid OSTAR2002 
(Lanz \& Hubeny \cite{ostar2002}), but for late spectral types, they are larger by 
up to 0.3 dex compared to the results of Smith et al.\ (\cite{smith})
based on WM-BASIC models. 
This shows that current atmosphere codes
are not fully consistent and indicates that the typical uncertainty on 
current ionising fluxes per unit area may still be up to a factor of $\sim$ 2 
for a given effective temperature. 

\end{itemize}

Our results should be tested further when more analysis
of individual stars are available and cover the whole range of
spectral types and luminosity classes.  
Despite some remaining uncertainties on the \teff\ scale, our
results should represent a significant improvement over previous calibrations, 
given the detailed treatment of non-LTE line-blanketing in the expanding atmospheres 
of massive stars. 

%%%%%%%%%%%%%%%%%%%%%%%%%%%%%%%%%%%%%%%%%%%%%%%%%%%%%%%%%%%%%%%%%%%%%%%%
\begin{acknowledgements}
We thank the referee, Rolf Kudritzki, for his suggestions and comments 
which contributed to improve this paper.
FM and DS acknowledge financial support from the Swiss National Found for 
Scientific research (FNRS). DJH would like to acknowledge partial support 
for this work from NASA grants NASA-LTSA NAG5-8211 and NAG5-1280.
We thank Daniel Pfenniger for giving us access to his PC cluster on
which the computations of CMFGEN models have been run. FM thanks Yves Revaz 
for assistance related to the cluster.

\end{acknowledgements}

%%%%%%%%%%%%%%%%%%%%%%%%%%%%%%%%%%%%%%%%%%%%%%%%%%%%%%%%%%%%%%%%%%%%%%%%
{}

%%%%%%%%%%%%%%%%%%%%%%%%%%%%%%%%%%%%%%%%%%%%%%%%%%%%%%%%%%%%%%%%%%%%%%%%
\end{document}